\documentclass[3p]{elsarticle}

\usepackage{siunitx}
\usepackage{graphicx}
\usepackage{booktabs}
\usepackage{amsmath}
\usepackage{float}
\usepackage{color}

\usepackage{mathtools}

\journal{Chemical Physics}

 \biboptions{ numbers,sort&compress}

\begin{document}

\begin{frontmatter}

\title{Coupled rate-equation hydrodynamic simulation of a Rydberg gas Gaussian ellipsoid:  Classical avalanche and evolution to molecular plasma }

\author{Rafael Haenel$^1$ and Edward Grant$^{*1,2}$}
\address{$^1$Department of Physics \& Astronomy, University of British Columbia, Vancouver, BC V6T 1Z1 Canada}
\address{$^2$Department of Chemistry, University of British Columbia, Vancouver, BC V6T 1Z1 Canada}


%\author{Edward Grant}\address{Department of Physics \& Astronomy, University of British Columbia, Vancouver, BC V6T 1Z1 Canada}
%\address{Department of Chemistry, University of British Columbia, Vancouver, BC V6T 1Z1 Canada}

%\author[mymainaddress,mysecondaryaddress]{Department of Chemistry, University of British Columbia, Vancouver, BC V6T 1Z1 Canada}

\cortext[mycorrespondingauthor]{Corresponding author}


\begin{abstract}

An ellipsoidal volume of Rydberg molecules, entrained in a supersonic molecular beam, evolves on a nanosecond timescale to form a strongly coupled ultracold plasma. We present coupled rate-equation simulations that model the underlying kinetic processes and molecular dissociation channels in both a uniformly distributed plasma and under the conditions dictated by our experimental geometry.  Simulations predict a fast electron-driven collisional avalanche to plasma followed by slow electron-ion recombination. Within 20 $\mu$s, release of Rydberg binding energy raises the electron temperature of a static plasma to $T_e = 100$ K. Providing for a quasi-self-similar expansion, the hot electron gas drives ion radial motion, reducing $T_e$.  These simulations provide a classical baseline model from which to consider quantum effects in the evolution of charge gradients and ambipolar forces in an experimental system undergoing responsive avalanche dynamics.

\end{abstract}

\begin{keyword}
Rydberg gas, avalanche, Penning ionization, electron impact, dissociative recombination, predissociation, nitric oxide

%\texttt{elsarticle.cls}\sep \LaTeX\sep Elsevier \sep template
%\MSC[2010] 00-01\sep  99-00
\end{keyword}

\end{frontmatter}

%\linenumbers

\section{Introduction} 

Ultracold Rydberg gases provide a very useful medium in which to study the dynamics of strong interactions in many-body quantum systems.  In a state of high principal quantum number, $n$, the highly polarizable charge distribution of a Rydberg atom produces a transient electric field that mixes the eigenstates of its neighbours \cite{Reinhard}, giving rise to interatomic forces \cite{Fioretti,Beguin} and, when prepared by a narrow bandwidth laser field, to the possibility of dipole blockade \cite{Comparat,Dudin}, coherent population trapping \cite{AdamsCoherPopTrap} and electromagnetic induced transparency (EIT) \cite{Gorshkov,Maghrebi,AdamsEIT}.  Such interactions produce states of entanglement that hold promise for the engineering of quantum gates \cite{SaffmanRMP,Gorniaczyk,Tiarks,AdamsQIP,Saffman} and exotic nonlinear optical materials \cite{Chang,Firstenberg}.

The complex many-body physics of Rydberg systems extend from ultra long-range dimers formed by pairwise scattering resonances \cite{Bendkowsky} to crystals \cite{BlochRydStr} and exotic phases in driven dissipative systems \cite{Carr,Gorshkov_disipative}.  Investigations in concept and in practice have pushed systems to regimes of high density in an effort to study Rydberg aggregates \cite{Pfau_nmer,Urvoy} and polaritons \cite{WeideManyBody,Bienias,Urvoy,PfauPolaritons}.  In a context of spin Hamiltonians, interacting Rydberg systems show promise for the quantum simulation of transport processes in condensed matter \cite{Browaeys,Noh}, or as engineered materials themselves with properties of metallic or dipolar crystalline solids \cite{Glaetzle,LiMetal,Lan}.  

Extreme states such as these form under conditions of excited state density in which inter-particle separations approach Rydberg orbital diameters.  Here, Penning interactions \cite{Robicheaux.2005} and interatomic Coulombic decay \cite{ICD} can produce ions and free electrons.  Kinetic electrons collide freely with Rydberg atoms, stimulating level redistribution, electron heating and avalanche to plasma.  This process has importance, both as an uncontrolled end state in the many-body physics of Rydberg systems, and as a complex system of significance in its own right.  

In particular, laboratory studies of strongly coupled plasmas provide a revealing means to probe the dynamics of many-body processes of relevance to a broad range of exotic but important charged-particle systems. Because classical Coulomb dynamics scale \cite{Scaling}, the properties of strongly coupled systems at low density connect directly to a very diverse range of phenomena from the evolution of stars and planets to low-temperature charge transport in complex materials. Energetic plasmas become strongly coupled at extremely high densities, when the average distance between ions, $a_{ws}$, nears the Bohr radius \cite{Ichimaru}. Ultracold plasmas approach states of strong coupling ($\Gamma = e^2/4\pi \epsilon_0a_{ws}k_BT > 1$) by virtue of their low temperatures. Created from ultracold atoms held in a magneto-optical trap (MOT), or molecules seeded at higher density in a supersonic expansion, plasmas with ion temperatures, $T_i$, of 1 K can reach values of $\Gamma_i$ from 3 to nearly 30 \cite{PhysRept}. Strongly coupled ultracold plasmas can manifest liquid-like or even crystalline properties. 

An ultracold plasma thus affords an opportunity to study strong coupling on a laboratory length scale.  Applying conventional spectroscopic methods, one can probe optically thin samples to gauge energy states and image density in three dimensions.  The short-time dynamics of avalanche from Rydberg gas to plasma appear to conform with kinetic-theory models based on particle-particle interactions \cite{PPR,Saquet:2012}.  Yet systems have shown evidence of self-organization to form spatial and temporal structures with fluid properties \cite{SchulzWeiling:2016}.  

We have developed a means of studying ultracold plasmas that starts by producing a state-selected Rydberg gas from a molecular precursor cooled to temperatures well below 1 K in a skimmed supersonic molecular beam \cite{Morrison:2008}.  Here, high, controlled ion density and neutral channels of dissociation add control points that offer a route to stronger coupling and perhaps new physics. 

Our experiment shows that spontaneous avalanche to plasma splits the core of an ellipsoidal Rydberg gas of nitric oxide \cite{SchulzWeiling:2016}.  Ambipolar expansion first quenches the electron temperature of this core plasma.  Then, long-range, resonant charge transfer from ballistic ions to frozen Rydberg molecules in the wings of the ellipsoid quenches the centre-of-mass ion/Rydberg molecule velocity distribution.  

This sequence of steps gives rise to a remarkable mechanics of self-assembly, in which the kinetic energy of initially formed hot electrons and ions drives an observed separation of plasma volumes.  These dynamics adiabatically sequester energy in a reservoir of mass transport, starting a process that anneals separating volumes to afford a state of cold, correlated ions, electrons and Rydberg molecules \cite{Haenel.2017}.  Short-time electron spectroscopy provides experimental evidence for complete ionization.  The long lifetime of this system, particularly its stability with respect to recombination and neutral dissociation, suggests that this transformation affords a robust state of arrested relaxation, far from thermal equilibrium.  We suggest that this state of the quenched ultracold plasma offers an experimental platform for studying quantum many-body physics of disordered systems.  

The semi-classical description of avalanche and relaxation in a high-density Rydberg gas forms a very important point of reference from which to interpret our experimental observations.  Coefficients validated by MD simulations serve in coupled rate-equation simulations to account well for the scaled avalanche dynamics of cold Rydberg gases, as long as they provide for a sufficient depth of electron-ion binding energy \cite{PVS, Niffenegger.2011,Scaling}.  We have developed models for plasmas of uniform density, in which we explore the effects of dissociative recombination and neutral predissociation on the evolution of electron temperature and lifetime \cite{Saquet:2011,Saquet:2012,Scaling}.  

The laser crossed molecular beam illumination volume creates a nonuniform Rydberg gas, distributed as a Gaussian ellipsoid.  The consequences of this initial condition appear in its evident effect on the evolution dynamics \cite{Haenel.2017}.  In the present paper, we develop a reference framework for coupled rate processes that explicitly reckon with the evolution of the kinetic and hydrodynamic processes over a realistic distribution of ion/electron and Rydberg densities.  We find no combination of initial conditions that conforms classically with the state of arrested relaxation observed experimentally.

\section{Experimental background}

Models considered in this work refer to experimental observations carried out in either of two differentially pumped skimmed supersonic molecular beam ultracold plasma spectrometers.  In a short flight path machine, a molecular beam of nitric oxide, seeded at 1:10 in 5 bar of helium, transits a perpendicular grid (G$_1$) to enter a field-free region, capped by a second grid (G$_2$) and multichannel plate (MCP) detector assembly.  The G$_2$-detector assembly is mounted on a carriage, the $z$ position of which is determined by a mechanical feed through.  The length of the region between G$_1$ and G$_2$ is adjustable in the propagation direction, from $z= 3$ to 200 mm.  In the second instrument, the molecular beam propagates a pre-configured, field-free distance of either 400 or 600 mm to impact a perpendicularly mounted 75 mm MCP detector equipped with a phosphor screen anode. 

A Nd:YAG pumped dye laser pulse crosses the molecular beam in the $x$ coordinate direction at a chosen distance 10 to 30 mm in $z$ beyond the skimmer.   This pulse ($\omega_1$) excites ground state NO to the $N' = 0$ level of the A $^2\Sigma^+$ vibrational ground state.  A second dye laser pulse ($\omega_2$) elevates this population to form state-selected Rydberg gas that propagates in $z$ as a Gaussian ellipsoid, 
\begin{equation}
\rho\left(x,y,z\right)
	=\rho_0 \exp \left( -\frac{x^2}{2 \sigma_x^2} -\frac{y^2}{2 \sigma_y^2} 
	-\frac{z^2}{2 \sigma_z^2} \right)
\label{eqn:geometry}
\end{equation}
with a well-defined velocity, longitudinal temperature ($T_{||}=500$ mK) transverse temperature ($T_{\perp}<5$ mK) and precisely known initial peak density in a range from $\rho_0=10^{10}$ to $10^{12}~{\rm cm^{-3}}$.  

In Eq. \ref{eqn:geometry}, $z$ refers to the propagation axis of the molecular beam, crossed by the laser beam along the axis $x$.  The diameter of the laser, measured in $y$ at the point of intersection, is about half that of the molecular beam.  For the purposes of the present simulations we assume a prolate Rydberg gas ellipsoid with an aspect ratio of 1.8:1.

A fraction of the Rydberg molecules undergoes Penning interactions to produce NO$^+$ ions and free electrons, as described below.  Electron-Rydberg collisions start an ionization avalanche that proceeds on a time scale from nanoseconds to microseconds depending on initial density and principal quantum number, $n_0$. 

Electrons gain energy from super-elastic collisions with Rydberg molecules, and the system evolves to a quasi-equilibrium of NO$^+$, e$^-$ and NO$^*$ in a distribution of high principal quantum numbers.  This relaxation and the transient state it produces conforms with the evolution from Rydberg gas to ultracold plasma observed in a magneto-optical trap (MOT) \cite{Li,WalzFlannigan}.  

Absorption of $\omega_2$ exclusively forms Rydberg states with total angular momentum neglecting spin $N=1$.  Only one such series has sufficient lifetime to form a plasma, the $N=1$ $f$ series converging to the ion rotational state, $N^+=2$, $nf(2)$.  We collect the spectrum of these transitions either by applying a ramped electrostatic field pulse to G$_1$, which yields the selective field ionization (SFI) spectrum of the Rydberg gas ensemble, or by simply waiting until the illuminated volume crosses the detection plane, which gives an electron signal proportional to the density of charge pairs.  The shape of this waveform describes the width of the excited volume in $z$ as a function of flight time, determined by the $z$-displacement of the detection plane. 

Selective field ionization (SFI) produces an electron signal waveform that varies with the amplitude of a linearly rising electrostatic field.  Electrons in a Rydberg state with principal quantum number, $n$, ionize diabatically when the field amplitude reaches the electron binding energy threshold, $1/9n^4$ in atomic units \cite{Gallagher}.  

For low density Rydberg gases, SFI has served as an exacting probe of the coupling of electron orbital angular momentum with core rotation.  Studies of nitric oxide in particular have shown that $nf(2)$ Rydberg states of NO traverse the Stark manifold to form NO$^+$ in rotational states $N^+=2$ and 0 \cite{Fielding}.  These experiments operate in a diabatic regime, employing a slew rate of 0.7 V cm$^{-1}$ ns$^{-1}$.  Under these conditions, SFI features that appear when the field rises to an amplitude of $F$ V cm$^{-1}$ measure electrons bound by energy $E_b$  in cm$^{-1}$, according to $E_b =4.12 \sqrt{F}$.  Quasi-free electrons, weakly bound in the attractive potential of more than one cation, ionize at a low field that varies with the number of excess ions in the plasma.  

\begin{figure}[h!]
    \centering
        \includegraphics[width=.4\columnwidth]{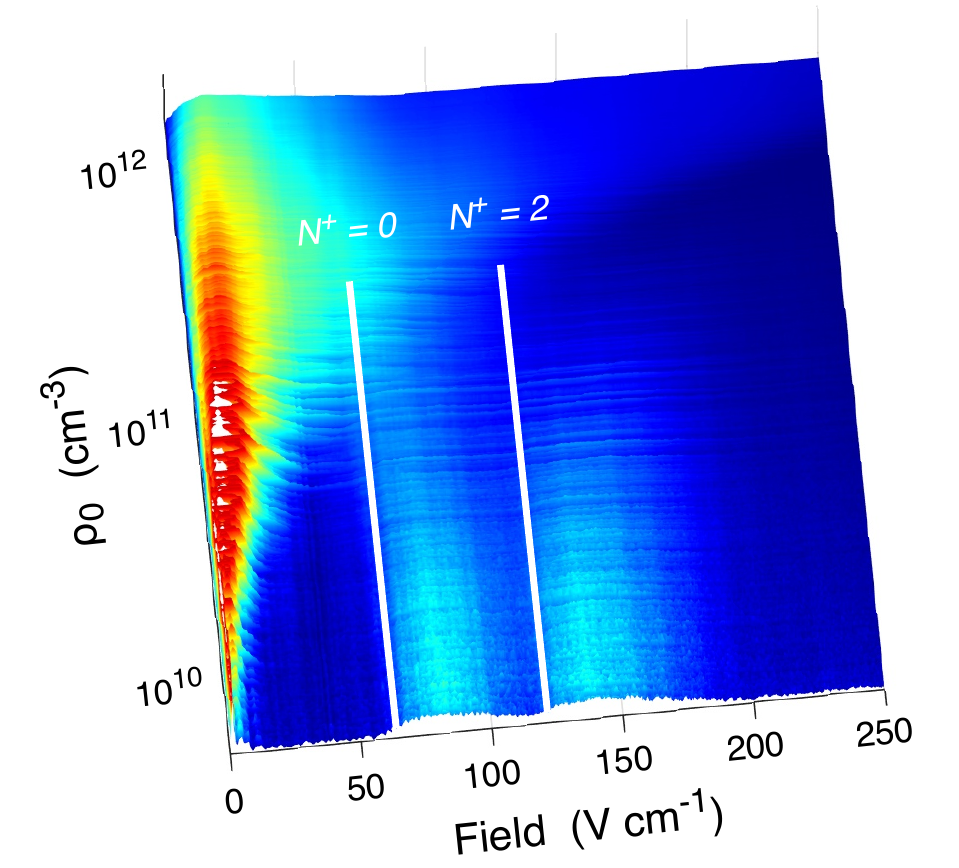}
    \caption {Selective field ionization spectrum as a function of initial Rydberg gas density, $\rho_0$, after 500 ns of evolution, showing the signal of weakly bound electrons ($0 - 20$ V cm$^{-1}$) combined with a residual population of $49f(2)$ Rydberg molecules.  These low-density features appear at 67 and 123 V cm$^{-1}$, corresponding respectively to the field-ionization thresholds to form NO$^+$ ion rotational states, N$^+$ = 0 and 2 from a Rydberg molecule with $n_0 = 49$ and a core rotational state, N$^+$ = 2.   The prominent feature that appears at the lowest values of the ramp field gauges the potential energy of electrons in high Rydberg states bound to single NO$^+$ ions, combined with electrons bound to the space charge of more than one ion.  The red feature extends approximately to the binding energy of $n_0=80$ or 500 GHz.}   \label{fig:SFI49}
\end{figure}

The SFI spectrum shown in Fig. \ref{fig:SFI49} maps the electron binding energy as a function of the initial Rydberg gas density for a molecular nitric oxide ultracold plasma after 500 ns of evolution.  The spectrum obtained at higher density  ($10^{12}~{\rm cm}^{-3}$) shows a signal appearing at low field that must represent quasi-free electrons bound to some space charge of excess NO$^+$ ions or Rydberg states of very high $n$.  

This contracts to a narrower distribution of very weakly bound electrons in plasmas of lower density ($10^{10}~{\rm cm}^{-3}$).  Here we also observe the spectrum of a residue of molecules with the originally selected principal quantum number of the Rydberg gas, shifted slightly to deeper apparent binding energy by evident $l$-mixing or slight relaxation in $n$. 

We have used SFI measurements like these to characterize the avalanche and evolution dynamics of a great many Rydberg gases of varying density and initial principal quantum number.  Relaxation times vary, but all of these spectra evolve to form the same final spectrum of weakly bound electrons with traces of residual Rydberg gas for systems of low initial density.   

\section{Methods of coupled rate-equation simulation} \label{sect:Methods}

\subsection{Penning Ionization} \label{Penning}

Laser excitation forms a Rydberg gas in which a local density $\rho$ defines a mean nearest neighbour distance, or Wigner-Seitz radius of $ a_{WS} =  \left(3/4 \pi \rho \right)^{1/3} $.  For simulations discussed below, an initial core density of $ \rho_0=0.5 \times 10^{12}$ cm$^{-3} $ yields an Erlang distribution of nearest neighbour separations with a mean value of $ 2 a_{WS}=1.6$  $\mu$m \cite{Torquato.1990}.  Modelling the Penning interaction semi-classically, Robicheaux \cite{Robicheaux.2005} predicts that 90 percent of Rydberg pairs within a critical distance of $ r_c = 1.8 \cdot 2 n_0^2 a_0 $ undergo Penning ionization within 800 Rydberg periods.  We determine the local density of Penning electrons ($ \rho_e$ at $t=0$), for any given initial local density, $\rho_0$, by integrating the Erlang distribution from $ r=0 $ to the critical distance $r = r_c$:
\begin{equation}
\rho_e(\rho_0,n_0) = \frac{0.9}{2} \cdot 4 \pi \rho_0 ^2\int_0^{r_{c}} r^2 \mathrm{e}^{-\frac{4\pi}{3}\rho_0 r^3}\mathrm{d}r \quad.
\label{eqn:Erlang}
\end{equation}
Here, the prefactor $ 0.9/2 $ reflects 90 percent ionization and accounts for the fact that two Rydberg molecules release only one Penning electron. Evaluating this definite integral yields an expression in closed form for the density of pairs that fall within this critical distance for a given initial density and principal quantum number,  
\begin{equation}
\rho_e(\rho_0,n_0) =\frac{0.9 \rho_0}{2}(1-\mathrm{e}^{-\frac{4\pi}{3}\rho_0 r_c^3}) \quad.
\label{Eq:PenDens}
\end{equation}
This defines the Penning electron density with which we initialize coupled rate-equation simulations, and $\rho_e(t=0)/ \rho_0$ determines the Penning fraction.  Energy conservation demands relaxation of the partner Rydberg to a quantum number lower than $ n_0/\sqrt{2} $. The semi-classical model predicts a level population proportional to $ n^5 $ \cite{Robicheaux.2005}.

\subsection{Coupled Rate-Equation Model for Uniform Density }

\subsubsection{Mechanism}

The plasma forms as the result of an avalanche of ionizing Rydberg-electron collisions described by a second-order rate constant $ k_i^{ion} $.
Three-body-recombination, with rate constant $ k_i^{tbr} $, reverses this process:
\begin{equation}
		\textrm{NO}^* + e^- \xrightleftharpoons[k^{tbr}]{k^{ion}} \textrm{NO}^+ + 2e^- \quad .
\label{ionization}\end{equation}
Both rate constants depend on electron temperature and vary with $n_i$, the principal quantum number of NO$^*$.  

Rydberg-electron collisions, $ k_{ij} $, scramble the initial state of NO$^*$, exchanging binding energy with electron thermal motion:
\begin{equation}
\textrm{NO}^*(n_i) + e^- \xrightleftharpoons[k_{ji}]{k_{ij}} \textrm{NO}^*(n_j) + e^- \quad.
\label{level}
\end{equation}

In a molecular Rydberg gas, dissociation affects plasma evolution. Predissociation of NO$^*$,
\begin{equation}
\textrm{NO}^* \xrightarrow[]{k^{PD}} \textrm{N}+\textrm{O} \quad ,
\label{PD}
\end{equation}
diminishes plasma density and consumes binding energy. Dissociative recombination,
\begin{equation}
\textrm{NO}^+ + e^- \xrightarrow[]{k^{DR}} \textrm{N}+\textrm{O} \quad ,
\label{p_DR}
\end{equation}
reduces the balanced population of electrons and ions. 

\subsubsection{Rate constants}

We adopt expressions for rate constants, $ k_i^{ion}$, $k_i^{tbr}$, $k_{ij} $ as functions of electron temperature developed with reference to Monte Carlo trajectory simulations by Pohl and coworkers \cite{Pohl.2008}.  Generally speaking, all of these rate constants increase as orbital radius grows with $ n $. High electron temperature favours ionization and suppresses recombination. Values of $ k_{ij} $ peak for transitions to neighbouring states. At high electron temperatures, $n$-changing collisions drive upward transitions more frequently than transitions down. Lower temperatures reverse this effect.  This relationship between the electron temperature and the net flux of Rydberg binding energy supports the formation of a quasi-equilibrium distribution of high-Rydberg molecules, ions and electrons at a temperature that depends on the density of the system.  Hydrodynamic expansion reduces electron temperature and thereby links the position of the quasi-equilibrium to evolution in the radial coordinates of the system.  

We calculate a rate constant for dissociative recombination $ k^{DR} $ based on the model of Schneider \textit{et al.} \cite{Schneider.2000}.  $ k^{DR}(T_e) $ fits a power-law in temperature with exponent $ {-0.51} $. Values of $ k^{PD} $ vary with $n$ and $l$ according to our previously introduced model \cite{Saquet.2012}.  In this picture, the increasing volume of Rydberg orbitals with principal quantum number decreases the predissociation rate by a factor of $ n^{-3} $.  The predissociation rate also depends on orbital momentum, $ l $.  For present purposes, we assume intrinsic rates $ k_l=0.014,0.046,0.029,0.0012 $ for $ l=0 $ to $ 3 $ and $ k_{l\ge 4} = \SI{3e-5}{} $ as compiled  by Gallagher \textit{et al} \cite{Murgu.2001} and Bixon and Jortner \cite{Bixon.1996}.  Electron collisions drive transitions between orbital momentum states, randomizing  the $ l $- and $ m_l $-distribution \cite{Chupka.1993}.  Statistical dilution further decreases $k^{PD}_i$ by a factor of $ n_i^{-2} $ to yield: 
\begin{equation}
k^{PD}_i=\frac{\sum_{l=0}^{n_i-1}(2 l +1) k_l}{n_i^2} \frac{\SI{4.13d7}{\per\nano\second}}{2 \pi n_i^3} \quad.
\label{eqn:kpd_model}
\end{equation}
For $ n=49 $, Eq. \ref{eqn:kpd_model} yields a lifetime $ \tau = (k^{PD}_{i=49})^{-1}=\SI{114}{\nano \second} $. Radiative decay is usually slower than predissociation and can thus be neglected \cite{Chupka.1993}.
Experimental lifetimes of $ nf(2) $-Rydberg states measured by Vrakking and Lee agree reasonably with this model \cite{Vrakking.1995}.

\subsubsection{Master equation for a uniform plasma density}

For a uniform density, $ \rho_i $,  of Rydberg molecules in state $ n_i $ and electron density, $ \rho_e $   the rate processes represented by Eqs. \ref{ionization} to \ref{p_DR} yield the system of coupled rate equations:
\begin{equation}
\begin{aligned}
	\frac{d \rho_i}{dt}  = -\rho_i \rho_e \sum_{j} k_{ij} + \rho_e \sum_{j} k_{ji} \rho_j 
	-k_i^{ion} \rho_i \rho_e 
	+ k_i^{tbr} \rho_e^3 - k_i^{PD} \rho_i\\
	\frac{d \rho_e}{dt}  = \rho_e \sum_{i} k_i^{ion}\rho_i - \rho_e^3 \sum_i  k_i^{tbr}   
	- k^{DR} \rho_e^2 \quad .
\end{aligned}
\label{eqn:rateeqn}
\end{equation}
where we assume quasi-neutrality ($ \rho_e=\rho_{ion} $).  Introducing the density of dissociated molecules $ \rho^{DR}$, $\rho_i^{PD} $ due to recombination and predissociation, respectively, readily confirms the conservation of particle number, $ N/V=\rho_e + \rho^{DR} + \sum_i \left( \rho_i + \rho_i^{PD} \right) $. 

In principal, the dynamics extend over an infinite range of quantum numbers $ n_i $.  In practice, however, we define a lower bound of $ n_{min}=10 $ and set the upper bound to $ n_{max} = \sqrt{R/k_B T} $, which conforms best with plasma scaling \cite{Saquet.2012}.

Energy conservation as an intensive variable in the uniform model determines the instantaneous electron temperature by, 
\begin{eqnarray}
0=\frac{d}{dt}E_{tot} = \frac{d}{dt} \left( \frac{3}{2} k_B T_e(t) \rho_e - R \sum_i \frac{\rho_i}{n_i^2} - E_{loss} \right) \quad,
\label{eqn:rateeqn2}
\end{eqnarray}
where $ E_{loss} $ is the energy loss per unit volume due to molecular dissociation:
\begin{eqnarray}
E_{loss} = - \frac{3}{2} k_B \int_{t_0}^{t} dt' T_e(t') \dot{\rho}_e^{DR}(t') + R \sum_i \frac{\rho_i^{PD}}{n_i^2} \,.
\label{eqn:rateeqn3}
\end{eqnarray}

The laser-crossed molecular beam geometry defines an illumination volume with a Gaussian ellipsoidal distribution (Eq. \ref{eqn:geometry}), which for present purposes has a peak initial density of  $ \rho_0 = \SI{0.5e12}{\centi\meter\tothe{-3}} $ and negligible Rydberg kinetic energy in the moving frame of the molecular beam.  To choose a representative uniform initial Rydberg gas density for use in the coupled rate equations above, we consider Gaussian ellipsoid surfaces of constant density $ S( r) $, which grow in area with principal axis distance as approximately $r^2$ and thus increase the portion of molecules in regions of lower density.  Numerically maximizing the expression $ S(r)\exp(-r^2/2\sigma^2) $, we find the most probable density $\rho_{rep} = 0.37\rho_0 $ and choose this quantity a most representative value for this reference model of an infinite, non-expanding plasma of uniform density.

 \subsection{Ellipsoidal Density Distribution}
As argued above, the Rydberg population does not sharply peak at $ \rho_{0} $. In fact, 50 percent of all Rydberg molecules fall in the initial density-interval $0.17\rho_0$ to $ 0.45\rho_0 $. To account explicitly for this Gaussian distribution of density, we construct an ellipsoid with semi-axes $ r_k = 5\sigma_k $ ($ k=x,y,z $), and divide this volume into $ L=100 $ ellipsoidal shells of equal width.

We assume a uniform particle density within each shell, which taken together, form a discretized density profile of the plasma. We then formulate separate sets of rate equations (Eq. \ref{eqn:rateeqn}) for each shell by introducing a new index $ \alpha=1,\dots,L $ that labels each shell:

\begin{equation}
\begin{aligned}
\frac{d \rho_i^\alpha}{dt}  = -\rho_i^\alpha \rho_e^\alpha \sum_{j} k_{ij} + \rho_e^\alpha \sum_{j} k_{ji} \rho_j^\alpha 
-k_i^{ion} \rho_i^\alpha \rho_e^\alpha + k_i^{tbr} \left(\rho_e^\alpha\right)^3 - k_i^{PD} \rho_i^\alpha\\
\frac{d \rho_e^\alpha}{dt}  = \rho_e^\alpha\sum_{i} k_i^{ion} \rho_i^\alpha - \left(\rho_e^\alpha\right)^3 \sum_i  k_i^{tbr}   
- k^{DR} \left(\rho_e^\alpha\right)^2 \quad .
\end{aligned}
\label{eqn:rateeqn_shell}
\end{equation}

At the density of the molecular beam, the recoiling neutral N($^4$S) + O($^3$P) products of nitric oxide dissociation have a mean free path that is much larger than the entire plasma volume \cite{SchulzWeiling.2016}.  The electrons on the other hand are bound by the space charge of the ions.  The volume occupied by the ions produced in a given shell define the evolving volume of that shell.  Thus, we assume that the constraint of quasi-neutrality confines the corresponding electrons to the same volume.  The low electron mass ensures a fast kinetic energy transfer between electrons in all shells, which we assume instantaneously equilibrates the electron temperature, $ T_e^\alpha=T_e^{\alpha'}=T_e $. As a consequence, energy conservation holds for the ellipsoid as a whole:
\begin{equation}
0=\frac{d}{dt} \sum_\alpha E_{tot}^\alpha(t)V^\alpha= \frac{d}{dt} \left( \frac{3}{2} k_B T_e(t) \sum_{\alpha} \rho_e^\alpha V^\alpha - R \sum_{i \alpha} \frac{\rho_i^\alpha}{n_i^2} V^\alpha - \sum_{\alpha}E_{loss}^\alpha V^\alpha \right) \,. \quad.
\label{eqn:energycons_shell}
\end{equation}
Equations \ref{eqn:rateeqn_shell}-\ref{eqn:energycons_shell} form a set of $ \left(n_{max}-n_{min}+2\right)L +1 $ ordinary differential equations which we solve by numerical integration.

\subsection{Hydrodynamic Shell Model}

Previous work has established the Vlasov equations as an effective means of describing the expansion of a spherical Gaussian plasma \cite{Dorozhkina.1998,PhysRept}.  We have introduced a shell model representation for the hydrodynamics of a Gaussian spherical plasma, and demonstrated that discrete forces determined by the gradient of the charge distribution at the boundary of each shell produce an equivalent self-similar expansion.  In the same work, we have extended this shell-model hydrodynamic approach to investigate the expansion hydrodynamics of shell-model density distributions flattened by dissociative recombination \cite{Sadeghi.2012}.

We have extended the analytical Vlasov equation approach to describe the collisionless expansion of a Gaussian ellipsoidal plasma and confirmed that these results match with an ellipsoidal shell-model numerical solution \cite{SchulzWeiling.2016}.   We are interested now to extend this approach to capture some of the effect of the loss of electron energy owing to expansion on the collisional dynamics of plasma avalanche and evolution, in other words, extend the shell model to include expansion effects. 

Under the conditions of simultaneous avalanche and expansion, the electron forces on the ions act to increase the radial coordinates of the plasma, diminishing its density and reduce the electron temperature according to:  
\begin{equation}
-m_i 	\frac{d u}{d t} =	e \nabla \Phi = k_B T_e \frac{\nabla \rho_e}{\rho_e} \quad,
\label{eqn:expansionforce_exact}
\end{equation}
where $ m_i $ represents the ion mass, and $ u $ the radial velocity.  These effects of electron temperature and charged particle density have straightforward effects on the coupled rate processes ongoing in the shells.  

But, composition changes in the shells owing to collisional processes cause the plasma density distribution to evolve in other than a self-similar expansion.  The high-density core of the ellipsoid ionizes first while most of the particles in the outer shells remain neutral.  The plasma thus begins life as a narrower Gaussian.  Driven by steeper gradients in the core, inner shells can overtake outer ones, inverting local density distributions and setting the stage for the formation of shock fronts.  However, higher-order collisional processes, including three-body and dissociative recombination, act to flatten charge-density distributions and thus moderate shockwave formation.    

The experiment shows striking evidence of plasma segmentation and bifurcation, but no shockwave singularities.  A more elaborate model for the charge gradients and local avalanche dynamics in the core and the wings of the ellipsoid could well explain the classical electrodynamics leading to gross changes in plasma morphology.  But, for present purposes we wish to determine how coupled rate processes, evolving at varying densities in the shells of a Gaussian ellipsoid respond to the global effects of plasma expansion and cooling owing to ambipolar expansion.  

We therefore simply neglect the effects of hydrodynamic expansion at the level of the shells, and enforce a self-similarity by approximating,
\begin{equation}
\label{eqn:expansionforce}
		-m_i \frac{d^2r_k^\alpha}{d t^2} \approx  k_B T_e \left< \frac{\nabla_k \rho_e}{\rho_e r_k} \right> r_k^\alpha = k_B T_e \frac{1}{L} \sum_{\beta} \frac{1}{\rho_e^\beta r_k^\beta} \frac{ \rho_e^{\beta+1}-\rho_e^\beta}{r_k^{\beta+1}-r_k^\beta} r_k^\alpha \quad.
\end{equation}
Here, $ r_k^\alpha $ is the position of shell $ \alpha $ along the $ k $-axis and brackets denote the average over all shells $ \alpha $. The second equality discretizes the gradient. 

Ambipolar expansion transfers electron kinetic energy to the radial motion of the ions.  For simplicity, we assume three uncoupled energy channels as would be the case for a spherical distribution:
\begin{equation*}
\label{eqn:E_kin}
		E_{kin}^\alpha \approx \frac{m_i}{2} \frac{ \left(u_x^\alpha\right)^2 + \left(u_y^\alpha\right)^2 + \left(u_z^\alpha\right)^2}{3} \quad .
\end{equation*}
Adding this to the energy conservation Eq. $ \ref{eqn:energycons_shell} $ gives:
\begin{equation}
0=\frac{d}{dt} \left( \frac{3}{2} k_B T_e(t) \sum_{\alpha} \rho_e^\alpha V^\alpha- R \sum_{i \alpha} \frac{\rho_i^\alpha}{n_i^2} V^\alpha - \sum_{\alpha}E_{loss}^\alpha V^\alpha + \sum_{\alpha}E_{kin}^\alpha V^\alpha \right) \,. 
\label{eqn:energycons_shell_hydro}
\end{equation}
Finally, we account for the rarefaction of the charge and Rydberg molecule distributions owing to expansion. Modifying Eq. \ref{eqn:rateeqn_shell} yields
\begin{equation}
\begin{aligned}
\frac{d \rho_i^\alpha}{dt}  = -\rho_i^\alpha \rho_e^\alpha \sum_{j} k_{ij} + \rho_e^\alpha \sum_{j} k_{ji} \rho_j^\alpha 
-k_i^{ion} \rho_i^\alpha \rho_e^\alpha  
+ k_i^{tbr} \left(\rho_e^\alpha\right)^3 - k_i^{PD} \rho_i^\alpha - \frac{\rho_i^\alpha}{V^\alpha}\frac{dV^\alpha}{dt}\\
\frac{d \rho_e^\alpha}{dt}  = \rho_e^\alpha\sum_{i} k_i^{ion} \rho_i^\alpha - \left(\rho_e^\alpha\right)^3 \sum_i  k_i^{tbr}   
- k^{DR} \left(\rho_e^\alpha\right)^2 - \frac{\rho_e^\alpha}{V^\alpha}\frac{dV^\alpha}{dt} \quad ,
\end{aligned}
\label{eqn:rateeqn_shell_hydro}
\end{equation}
where the volume of shell $\alpha$ changes as 
\begin{equation}
	\frac{dV^\alpha}{dt}= \frac{4 \pi}{3} \frac{d}{dt}\left( \prod_k r_k^{\alpha} -\prod_k r_k^{\alpha-1} \right) \quad.
\label{eqn:volume}
\end{equation}
We may now proceed to integrate Eq. \ref{eqn:expansionforce}-\ref{eqn:volume}.

The following sections present simulation results and discussion for an $ n_0=49 $ Rydberg gas formed by double-resonant excitation at $t=0$ with a moderate density of $ \rho_0=\SI{0.5e12}{\centi\meter\tothe{-3}} $.  Here, the initial ellipsoid has Gaussian widths (Eq. \ref{eqn:geometry}) $ \sigma_x= \SI{0.75}{\milli \meter} $ and $ \sigma_y =\sigma_z= \SI{0.42}{\milli \meter} $ determined by the overlap of the laser and molecular beams.  Penning ionization provides the initial seed of free electrons according to Eq. \ref{Eq:PenDens} ($cf.$ Figure \ref{fig:penning})

We also examine the classical evolution of a lower-density plasma configured to match the experimental system observed after $ \SI{10}{\micro\second} $ of evolution. Here, we approximate a larger volume that has just begun to bifurcate by a pair of ellipsoids, of dimensions $\sigma_x = 1.0$ mm, $\sigma_y = 0.55$ mm and $\sigma_z = 0.70$, with either of two initial conditions supported by electron binding energy distributions seen in selective field ionization experiments:  (a) A Rydberg gas of high principal quantum number as represented by $n_0=80$, and (b) an ionized plasma with an initial temperature of $ T_e=\SI{5}{\kelvin} $.  Experimentally, we observe a plasma with such properties to exhibit arrested relaxation.  Coupled rate-equation models seek to test these models for classical stability. 

\section{Results and Discussion}

\subsection{Early-time evolution}

Coupled rate equation simulations begin with sets of initial conditions that define the selected state of the system at $t=0$.  For present purposes, we consider initial states that consist of a Rydberg gas of NO$^*$ with initial principal quantum number, $n_0$, or a plasma of NO$^+$ ions and free electrons with temperatures, $T_i$ and $T_e$. Calculations reported here assume an ion temperature of 0 K.  As noted above, we describe the local densities of ions and Rydberg molecules in terms of uniform distributions of well-defined initial density in the plasma as a whole or in each of the 100 layers of a shell-model Gaussian ellipsoid.  

In Rydberg gases, we assume that prompt Penning interactions between molecules in the leading edge of the nearest-neighbour distribution convert Rydberg molecules to form ion and electron fractions determined by the initial principal quantum number and local density as described above.  In the uniform Rydberg gas, this produces a well-defined initial density of Penning ions and electrons.  
\begin{figure}[h!]
\begin{center}
	\includegraphics[width=0.4\textwidth]{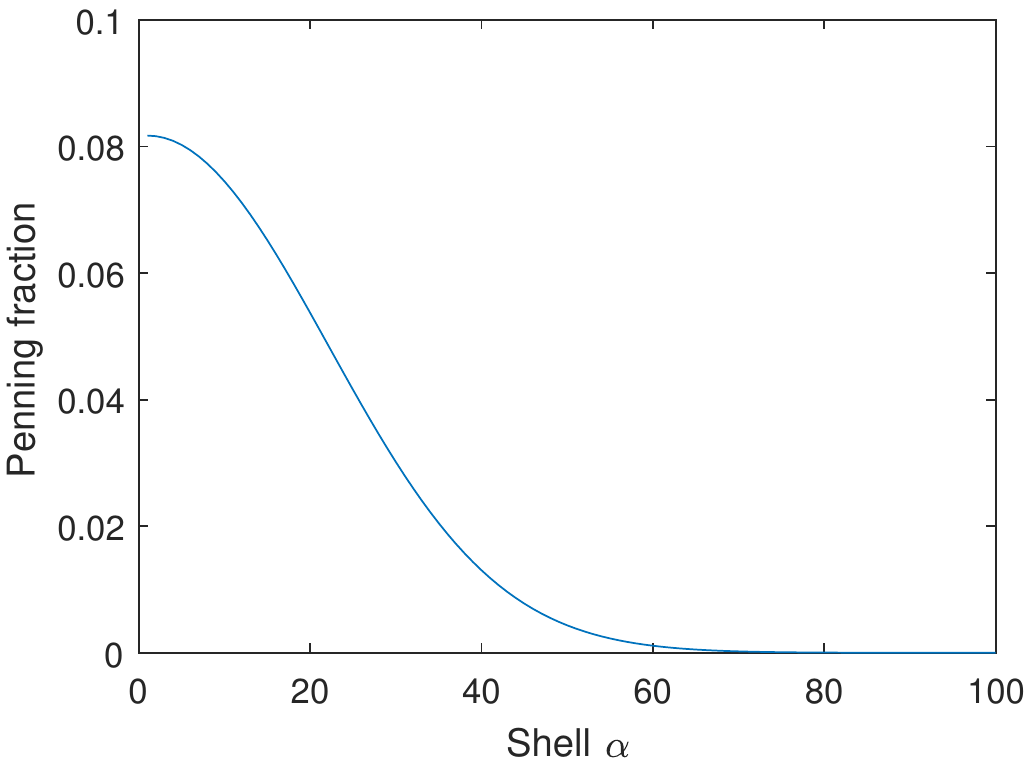}
	\caption{\label{fig:penning} Penning fraction as a function of shell number for a model Gaussian ellipsoid with dimensions, $\sigma_x = 0.75$ mm, $\sigma_y = \sigma_z = 0.42$ mm, and initial peak density of $0.5 \times 10^{12}$ cm$^{-3}$, represented by 100 shells spanning the interval from $r_{k} = 0$ to $5\sigma_{k}$ ($ k=x,y,z $), as determined by Eq. \ref{Eq:PenDens}.}
	\end{center}
\end{figure}
In the shell model, distinct nearest-neighbour interactions yield particular $t=0$ Penning densities for each shell.  For the initial conditions defined by a Gaussian ellipsoid with principal dimensions, $\sigma_x = 0.75$ mm, $\sigma_y = \sigma_z = 0.42$ mm, as defined by an intersection of the laser with the molecular beam, the Penning fraction in each shell varies almost linearly with its density.  Over the 100 shells constructed to describe the ellipsoid over radial distances of 5$\sigma$, the Penning fraction of initial NO$^+$ ions and electrons falls off as the approximate Gaussian displayed in Figure \ref{fig:penning}.  Systems configured initially evolve in time according to Eqs. \ref{eqn:rateeqn} to \ref{eqn:rateeqn3}.  Free electrons present initially collide frequently with stationary Rydberg molecules and ions.

\begin{figure}[H]
\begin{center}
	\includegraphics[width=.9\textwidth]{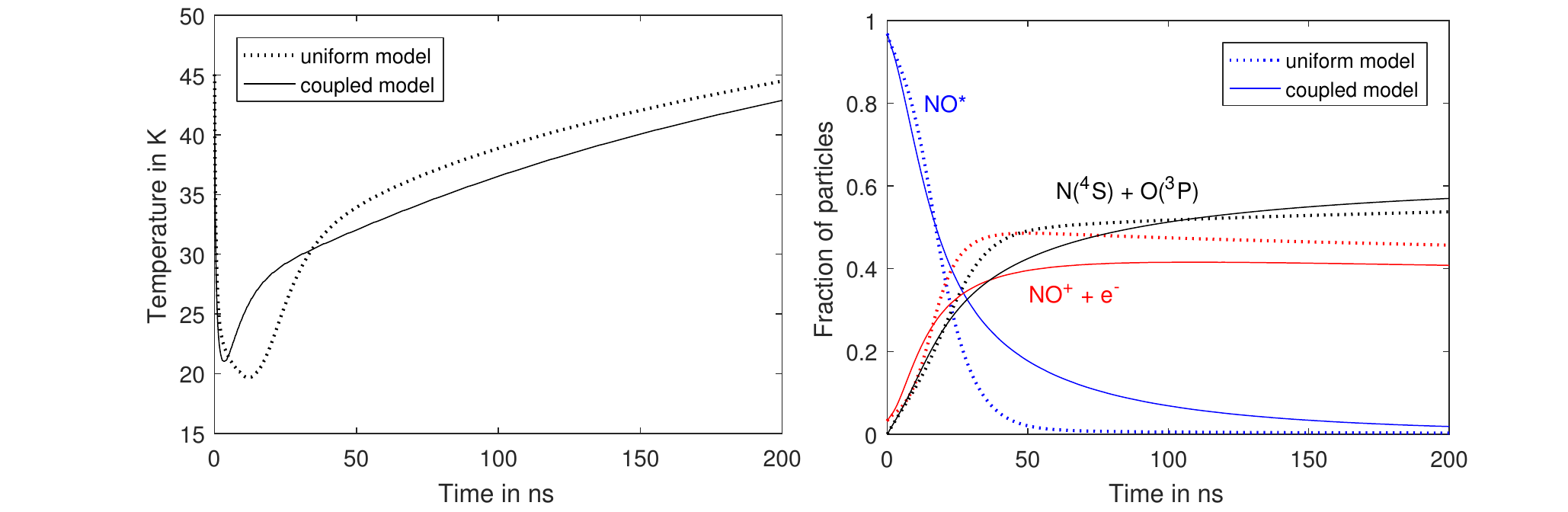}
	\caption{\label{fig:uniform_coupled} (left) The short-time evolution of electron temperature in plasmas formed by avalanche from a Rydberg gas with $n_0=49$ in an ellipsoid with a uniform initial density of $0.18 \times 10^{12}$ cm$^{-3}$ (dotted line) and a shell-model Gaussian ellipsoid with initial $\sigma_x = 0.75$ mm, $\sigma_y = \sigma_z = 0.42$ mm and an initial peak density of $0.5 \times 10^{12}$ cm$^{-3}$ (solid line).  (right) Global population fractions of particles as they evolve in the avalanche of model plasmas formed in a Rydberg gas ellipsoid with uniform density (dotted line) and the shell-model ellipsoid (solid line)  as above. }
\end{center}
\end{figure}

In a Rydberg gas, this starts an electron impact avalanche, which consumes NO$^*$ and drives the system to a quasi-equilibrium.   Figure \ref{fig:uniform_coupled} compares the dynamics of avalanche and electron heating that occurs in Rydberg gases defined by a uniform density distribution and an ellipsoidal shell model.  Note for comparable average densities, both systems signal the onset of avalanche with an initial electron temperature drop, followed by electron heating.  The interval of electron cooling develops more slowly in the uniform system, reaching a minimum over an interval about four times longer than the cooling phase of avalanche in the shell-model ellipsoid.  Thereafter, both systems rise in electron temperature at about the same rates.  
	
	In the uniform model, initial avalanche, driven by collisions of Penning electrons with abundant Rydberg molecules consumes electron kinetic energy. The temperature drops sharply to $  \SI{20}{\kelvin}$.  Rydberg molecules cascading to lower principal quantum number reverse this process by releasing their binding energy. This energy release increases in rate as the number of free electrons grows.  Eventually, the increased electron temperature shuts off three-body recombination.  Predissociation removes the reservoir of Rydberg binding energy and the system reaches a quasi steady state.

	In the ellipsoidal shell model the Rydberg gas evolves in much the same way. However, higher initial Penning fractions and an overall higher density in the core lead to a faster evolution of both temperature and avalanche in the first $ \SI{20}{\nano\second} $.  But as avalanche in the core finishes, the effects of the low density wings of the Gaussian model become more apparent. Here, small Penning fractions and dilute densities cause a very slow ionization process and lower temperatures. Predissociation does not depend on density. If the avalanche time exceeds predissociation lifetimes, Rydberg molecules simply dissociate before they ionize. The equilibration of electron temperature between the core and the cooler wings diminishes the longer term temperature rise in the ellipsoid.  

Note that cooling leads the global production of ions and electrons in the shell model, while lagging it to a slight degree in the system of uniform density.  Whether describing the local Penning electron density by a uniform representative value or by the Gaussian distribution of Figure  \ref{fig:penning}, the avalanching plasma forms neutral dissociation products with comparable initial rates, rising to completion faster in the case of the uniform plasma as can be observed in the right panel of Fig. \ref{fig:uniform_coupled}. 

\begin{figure}[h!]
\begin{center}
	\includegraphics[width=.34\textwidth]{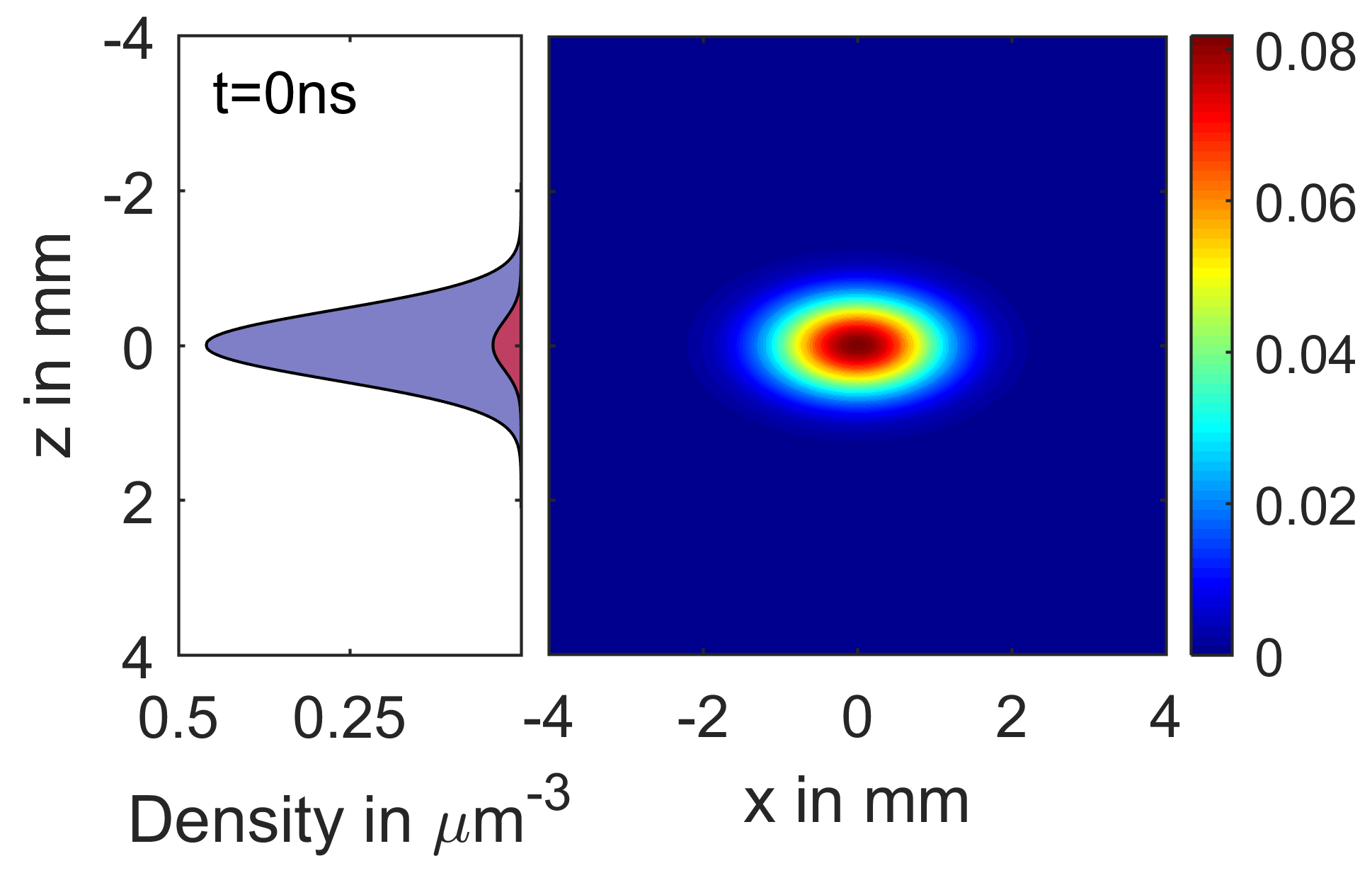}
		\includegraphics[width=.34\textwidth]{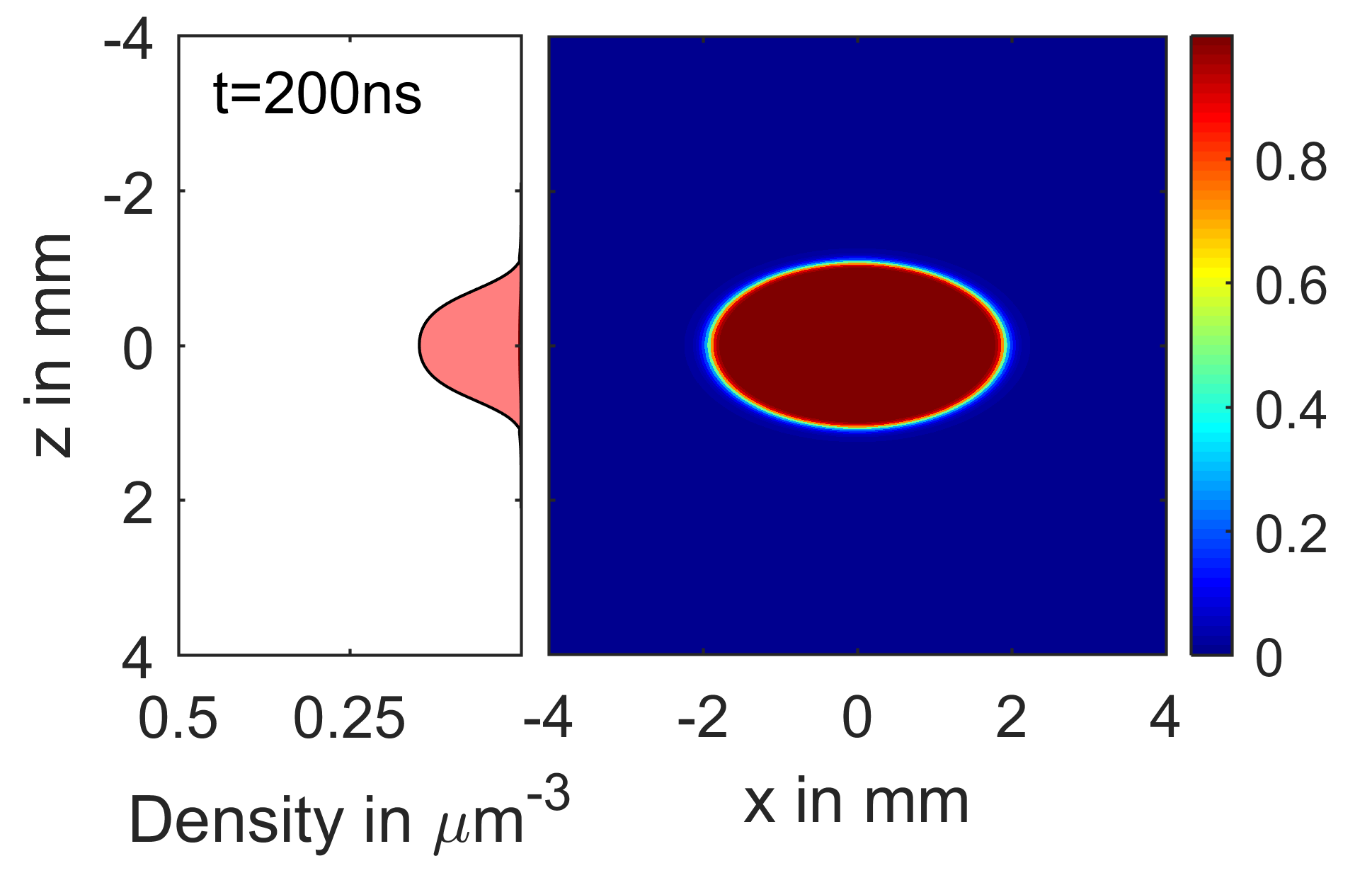}
	\includegraphics[width=.34\textwidth]{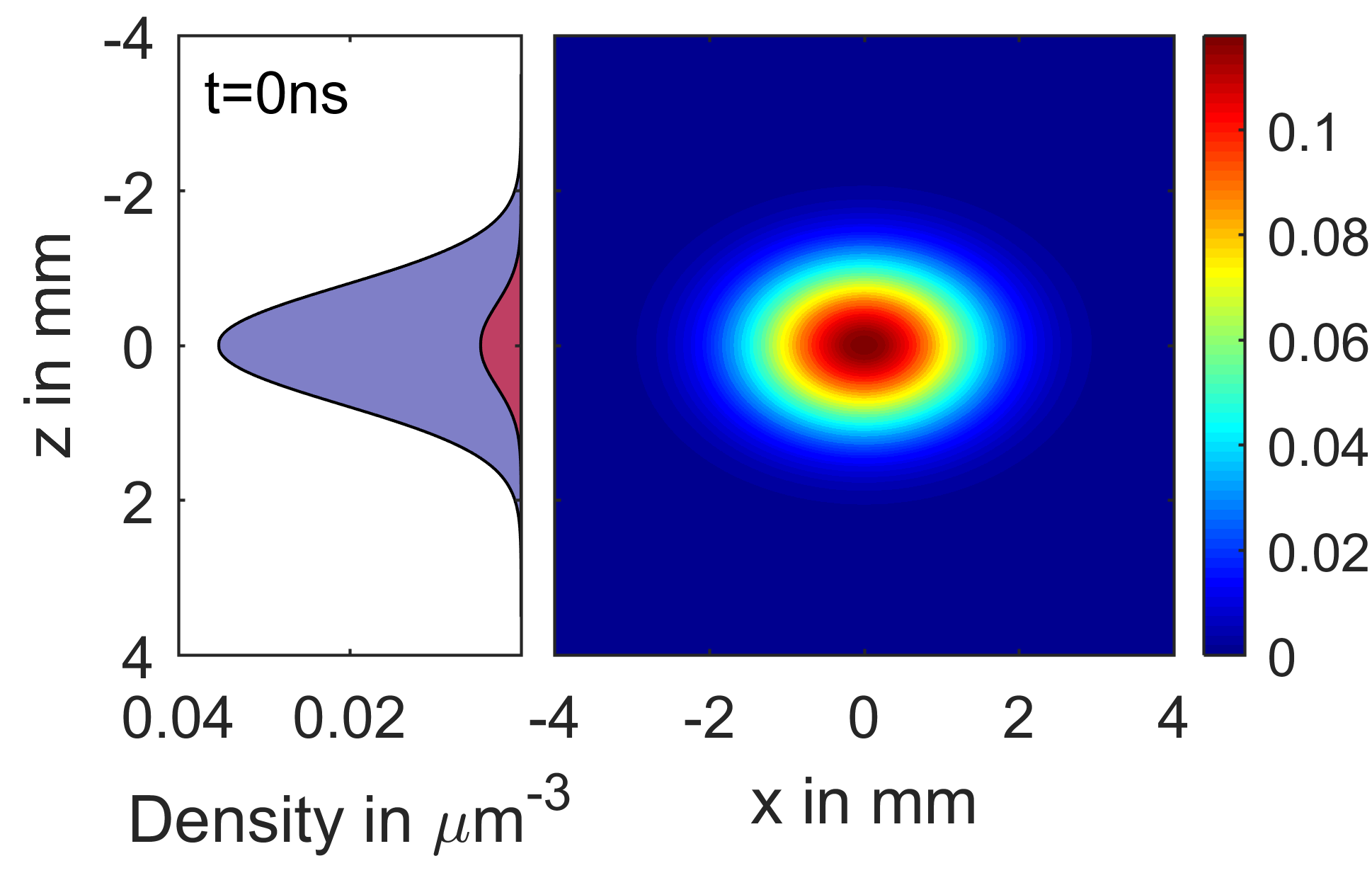}
	\includegraphics[width=.34\textwidth]{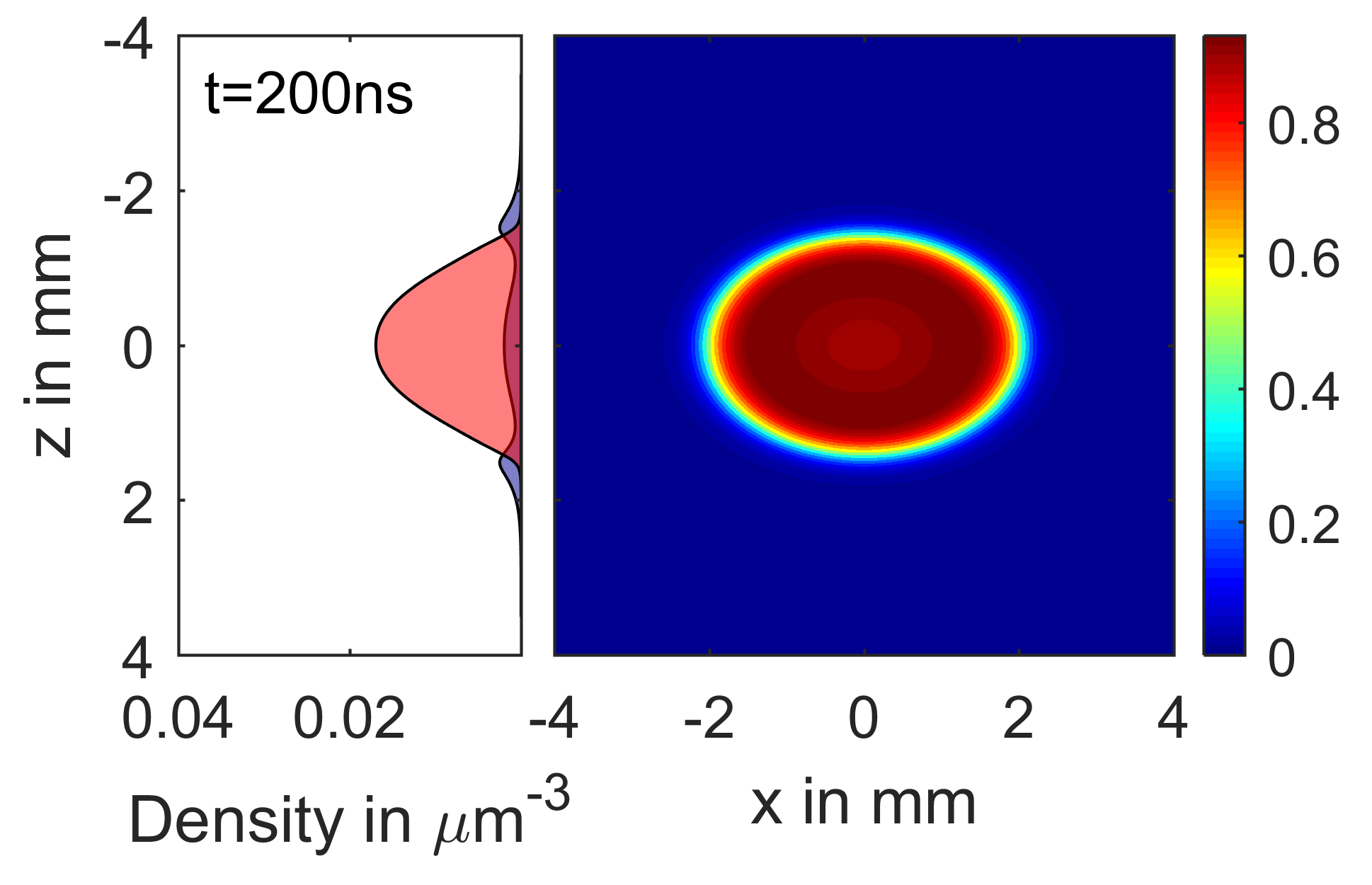} 
	\caption{\label{fig:shell_dist} Schematic model plasma volumes representing: (top left) The $\sigma_x = 0.75$ mm, $\sigma_y = \sigma_z = 0.42$ mm $n_0=49$ Rydberg gas ellipsoid formed with an initial peak density of $0.5 \times 10^{12}$ cm$^{-3}$ by the intersection of the $\omega_1$ laser pulse with a skimmed supersonic molecular beam. (top right) The same volume after a coupled rate equation simulation of 200 ns in the progress of  the relaxation and electron-impact avalanche.  (lower left) Idealized representation configured at $t=0$ to represent the experimental plasma observed after having evolved 10 $\mu$s to a state of arrested relaxation.  Here, we see an expanded ellipsoid ($\sigma_x = 1.0$ mm, $\sigma_y = 0.55$ mm and $\sigma_z = 0.70$ mm) of Rydberg gas quenched to a hypothetical state with a representative quantum number of $n_0=80$ and peak density of $0.4 \times 10^{11}$ cm$^{-3}$.   (lower right) The same volume after a coupled rate equation simulation of 200 ns, showing the predicted effects of relaxation and electron-impact avalanche.  Scale bar colour denotes the ratio of free electrons (and  NO$^+$ ions) to NO$^* $ Rydberg molecules.  Panels at the far left show Gaussian distributions of Rydberg gas, together with ion-electron densities at $t=0$ for both simulations based on the initial Penning fraction, calculated according to the method described in Section \ref{Penning}. }
	\end{center}
\end{figure}

We see this difference in the rate of decay of NO$^*$:  In both cases, Rydberg molecules decay faster than the rate of charged particle formation owing to the additional loss to neutral particles, but the global population of NO$^*$ survives a little longer in the shell-model ellipsoid. 
This occurs because significant numbers of Rydberg molecules in the wings fail to avalanche before they slowly predissociate. 

The distribution of mass in a realistic Gaussian ellipsoid causes the particle densities to evolve differently in each shell.  The top row of Figure \ref{fig:shell_dist} maps the ratio of free electrons to NO$^*$ Rydberg molecules, $\rho_e/\sum \rho_i$, for the $\sigma_x = 0.75$ mm, $\sigma_y = \sigma_z = 0.42$ ellipsoid at $t=0$ (as represented by Figure \ref{fig:penning}), and after an evolution of 200 ns.  

Here we see at $t=0$ that a very small fraction of the Rydberg gas forms Penning electrons and ions in the core of the ellipsoid.  These prompt Penning electrons seed an avalanche that in 200 ns proceeds nearly to completion in the inner 50 shells.  The low Rydberg density in the outer shells suppresses Penning ionization and retards avalanche.  These outer shells have low densities.  But they represent a very large portion of the volume of the ellipsoid, and thus figure significantly in determining the global numbers of Rydberg molecules compared with ions and electrons - the extent of ionization. 

Selective field ionization spectra establish that, after 10 $\mu$s, the plasma forms an arrest state with a narrow distribution of electron binding energies extending only 500 GHz below the ionization threshold of NO.  To represent the classical evolution of such a system using coupled rate simulations, we might choose initial conditions represented either by a Rydberg gas with $n_0=80$ or a fully ionized plasma with low-energy electrons ($T_e=5$ K), bound by a slight excess of plasma ions.  

For reference, Figure \ref{fig:shell_dist} begins a shell-model coupled rate equation simulation with $t=0$ initial conditions chosen to represent the experimental plasma at the point of arrested relaxation as an ellipsoid of Rydberg gas  ($\sigma_x = 1.0$ mm, $\sigma_y = 0.55$ mm and $\sigma_z = 0.70$ mm) quenched to a hypothetical state with an initial peak density of $0.4 \times 10^{11}$ cm$^{-3}$ and an initial principal quantum number of $n_0=80$.  Note after a further evolution of only 200 ns that rapid relaxation in the denser inner shells produces a plasma that looks completely ionized.  However, with volume scaling as the product of the three radial distances, even just a few unavalanched outer shells of low density hold a substantial number of Rydberg molecules, affecting the global balance between NO$^+$ and NO$^*$, as well as representing a residual heat capacity that discernibly affects the evolution of temperature.    

\begin{figure}[h!]
\begin{center}
	\includegraphics[width=.8\textwidth]{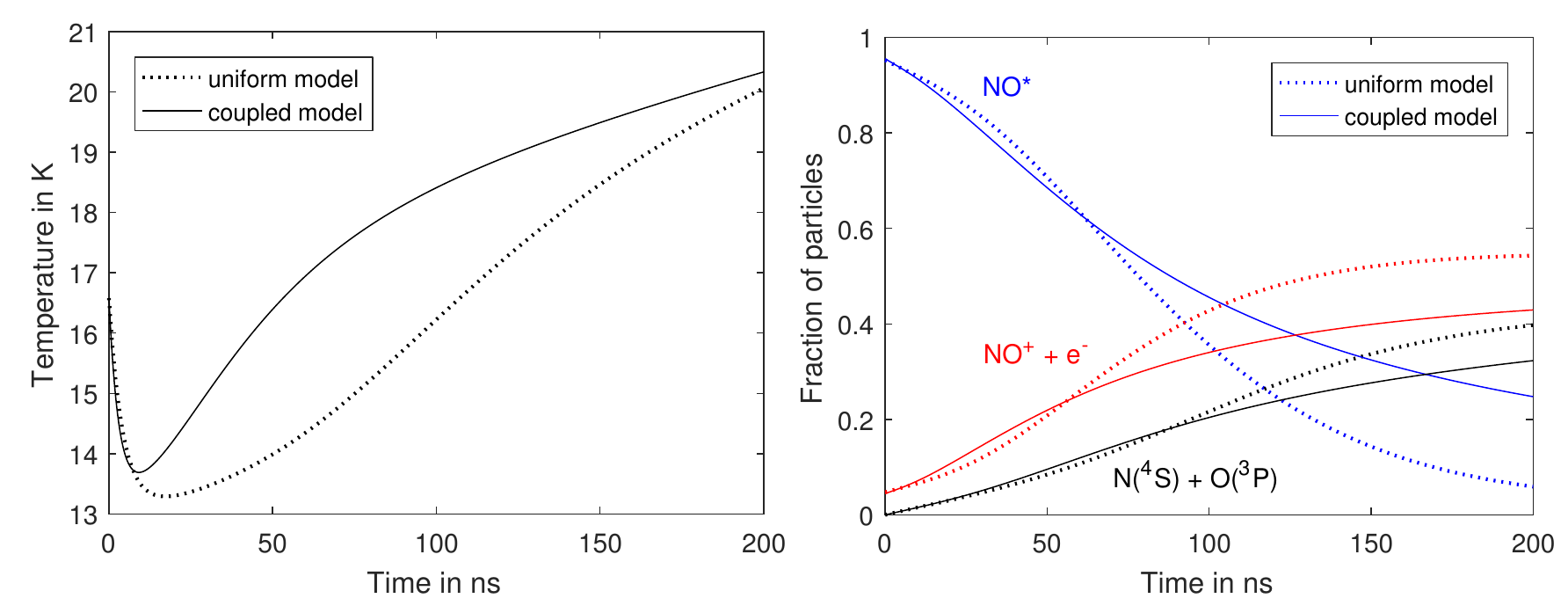}
	\caption{\label{fig:n80} (left) The short-time evolution of electron temperature in plasmas formed by avalanche from a Rydberg gas with $n_0=80$ in an ellipsoid with a uniform initial density of $0.14 \times 10^{11}$ cm$^{-3}$ (dotted line) and a shell-model Gaussian ellipsoid with $\sigma_x = 1.0$ mm, $\sigma_y = 0.55$ mm and $\sigma_z = 0.70$ mm and an initial peak density of $0.4 \times 10^{11}$ cm$^{-3}$ (solid line).  (right) Global population fractions of particles as they evolve in the avalanche of model plasmas formed in a Rydberg gas of uniform density (dotted line) and a shell-model ellipsoid (solid line) as above.  }
	\end{center}
\end{figure}

Figure \ref{fig:n80} traces the evolution of electron temperature and global particle densities for the Gaussian shell-model for an $n_0=80$ quenched plasma state compared with a uniform plasma of representative density.  We see even at low density that avalanche in either representation approaches the quasi-steady state in 200 ns.  Here we see how the increased cross-section of a higher Rydberg state compensates for the lower density.  The distribution of particle densities over a Gaussian ellipsoid of shells produces only small differences in the global evolution of particle densities and electron temperature. 
\begin{figure}[h!]
\begin{center}
	\includegraphics[width=.8\textwidth]{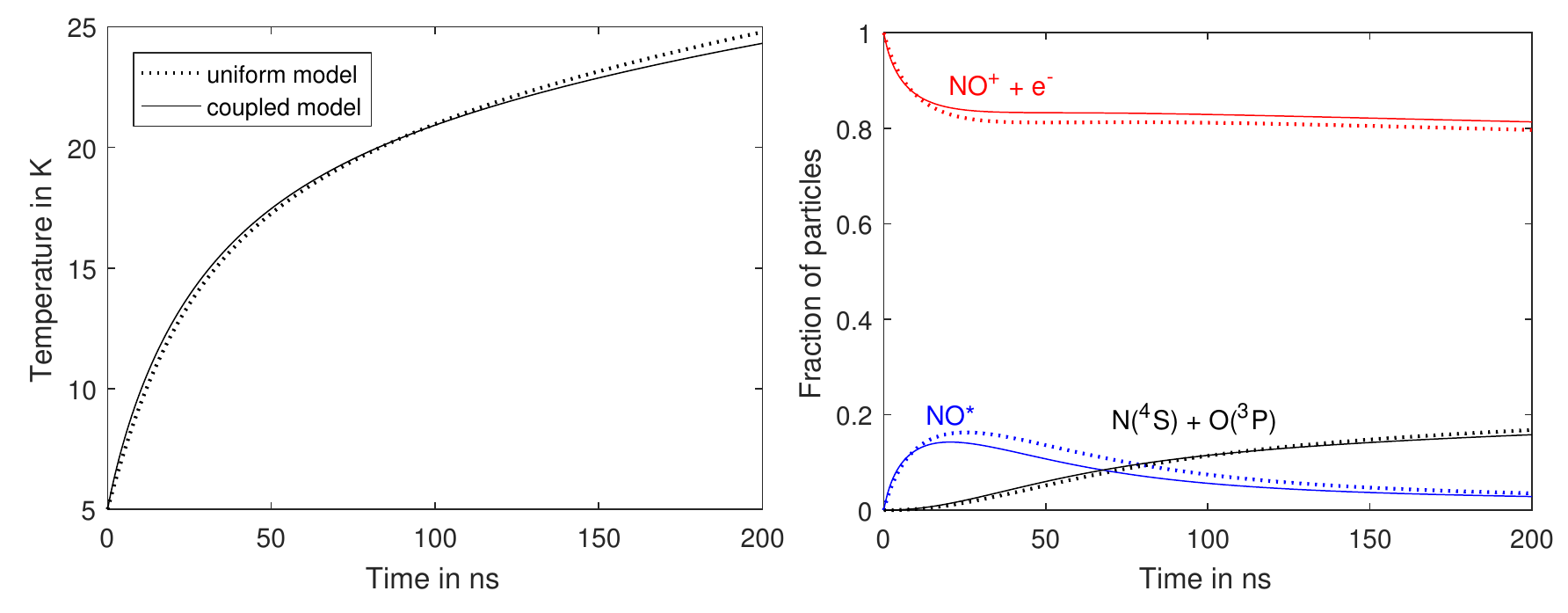}
	\caption{\label{fig:T5K} (left) The short-time evolution of electron temperature in an ultracold plasma with initial $T_e(0)=5$K in an ellipsoid with a uniform initial density of $0.14 \times 10^{11}$ cm$^{-3}$ (dotted line) and a shell-model Gaussian ellipsoid with initial $\sigma_x = 1.0$ mm, $\sigma_y = 0.55$ mm and $\sigma_z = 0.70$ mm and an initial peak density of $0.4 \times 10^{11}$ cm$^{-3}$ (solid line).  (right) Global population fractions of particles as they evolve in the avalanche of model plasmas formed of uniform density (dotted line) and a shell-model ellipsoid (solid line) as above. }
	\end{center}
\end{figure}

Figure \ref{fig:T5K} shows the results of a coupled rate simulation describing classical evolution when we choose the opposite limit to describe the quenched plasma state - that of a fully ionized gas with an initial electron temperature, $T_e=5$ K with a Gaussian ellipsoid of NO$^+$ ions with initial $\sigma_x = 1.0$ mm, $\sigma_y = 0.55$ mm and $\sigma_z = 0.70$ mm and an initial peak density of $0.4 \times 10^{11}$ cm$^{-3}$ described by 100 shells extending from 0 to 5$\sigma$.  Comparing with Figure \ref{fig:n80}, we see that in 200 ns the electron temperature rises to a similar degree and that the global evolution of particle densities differs very little in a Gaussian ellipsoid plasma as opposed to one described in terms of a representative uniform density. 

Both models call for substantial predissociation and evolution to a residual gas of ions and hot electrons.  Thus, we must conclude that neither the coupled rate-equation model of a Rydberg gas quenched to a state of $n_0=80$, or a fully ionized gas quenched to an electron temperature, $T_e=5$ K, serves to describe the experimentally observed short-time dynamics of the ultracold plasma state of arrested relaxation.  

\subsection{Long-time evolution}

\subsubsection{Static models}

\begin{figure}[h!]
\begin{center}
	\includegraphics[width= .9 \textwidth]{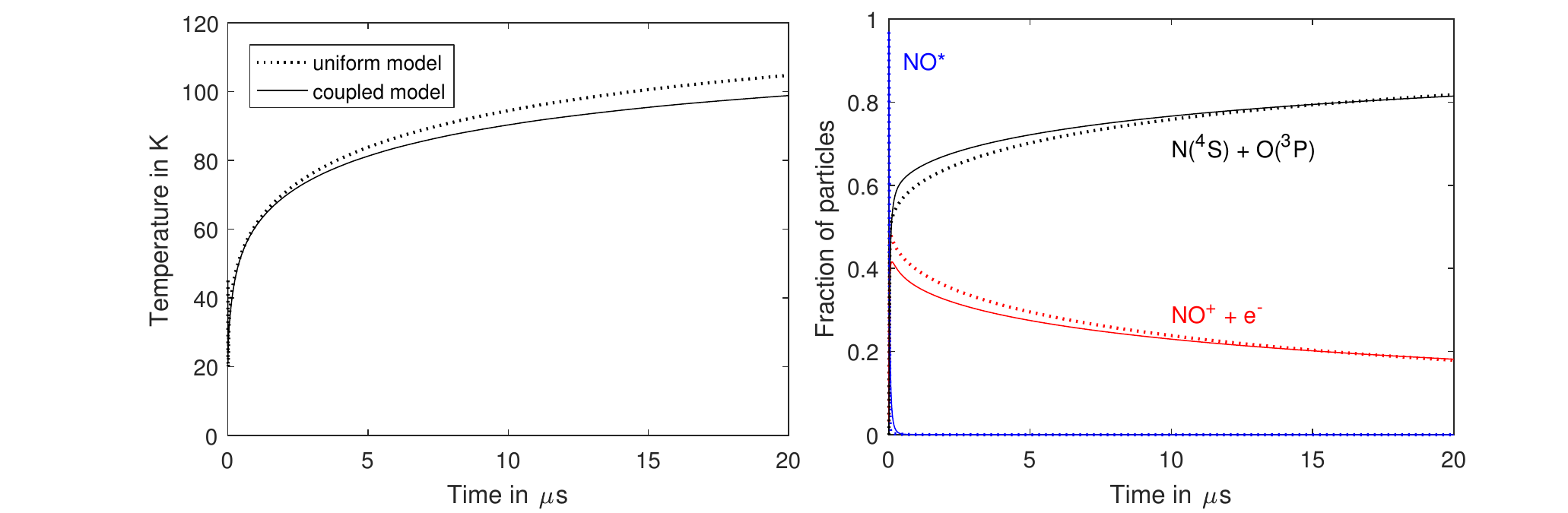}
	\caption{\label{fig:longtime49} (left) The long-time evolution of electron temperature in plasmas formed by avalanche from a Rydberg gas with $n_0=49$ in an ellipsoid with a uniform initial density of $0.18 \times 10^{12}$ cm$^{-3}$ (dotted line) and a shell-model Gaussian ellipsoid with initial $\sigma_x = 0.75$ mm, $\sigma_y = \sigma_z = 0.42$ mm and an initial peak density of $0.5 \times 10^{12}$ cm$^{-3}$ (solid line).  (right) Global population fractions of particles as they evolve in the avalanche of model plasmas formed in a Rydberg gas of uniform density (dotted line) and a shell-model ellipsoid (solid line) as above.  }
	\end{center}
\end{figure}

We turn now to the results of coupled-rate equation simulations of plasma evolution over the longer time interval of 20 $\mu$s.  Figure \ref{fig:longtime49} shows that, in terms of global properties, avalanche in a shell-model Gaussian ellipsoid with $\sigma_x = 0.75$ mm, $\sigma_y = \sigma_z = 0.42$ mm and an initial peak density of $0.5 \times 10^{12}$ cm$^{-3}$ proceeds very much like that in a system of representative uniform density.  Rydberg relaxation heats the gas of free electrons to a temperature of about 100 K.  As found for classical evolution on a 200 ns timescale, the initial distribution of NO$^*$ molecules and NO$^+$/e$^-$ charge pairs has little effect on the evolution of global energy and particle densities.  

In both cases predissociation fully depletes the Rydberg population before $ \SI{1}{\micro \second} $ leaving no molecules to ionize. Consequently, the ions and electrons undergo a second-order decay driven by dissociative recombination, which converts electrons into N($^4$S)+O($^3$P). 

\begin{figure}[h!]
\begin{center}
	\includegraphics[width= .9 \textwidth]{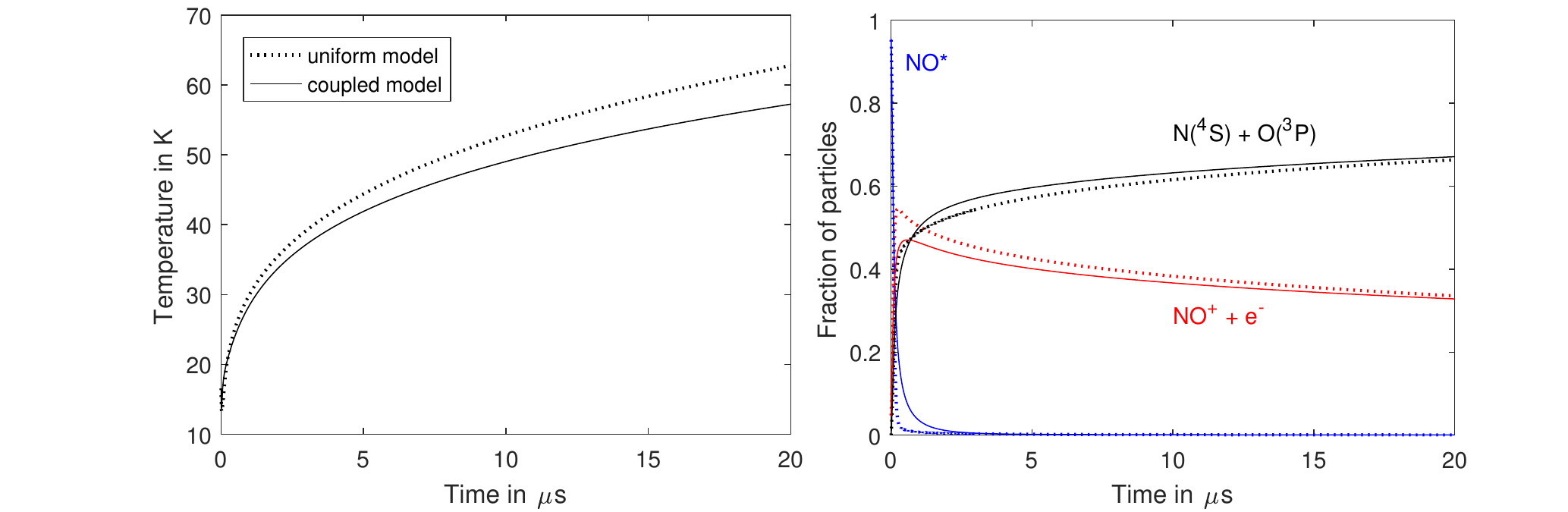}
	\caption{\label{fig:longtime80} (left) The long-time evolution of electron temperature in plasmas formed by avalanche from a Rydberg gas with $n_0=80$ in an ellipsoid with a uniform initial density of $0.14 \times 10^{11}$ cm$^{-3}$ (dotted line) and a shell-model Gaussian ellipsoid with initial $\sigma_x = 1.0$ mm, $\sigma_y = 0.55$ mm and $\sigma_z = 0.70$ mm and an initial peak density of $0.4 \times 10^{11}$ cm$^{-3}$ (solid line).  (right) Global population fractions of particles as they evolve in the avalanche of model plasmas formed in a Rydberg gas of uniform density (dotted line) and a shell-model ellipsoid (solid line) as above.  }
	\end{center}
\end{figure}

\begin{figure}[h!]
\begin{center}
	\includegraphics[width= .9 \textwidth]{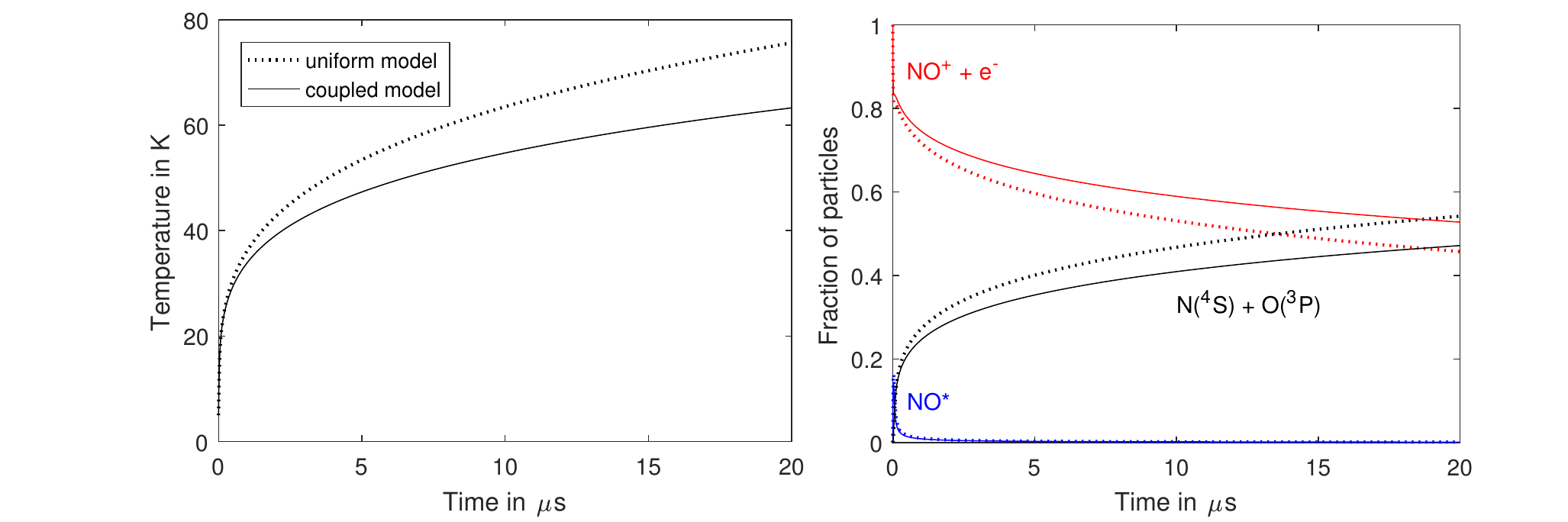}
	\caption{\label{fig:longtimeT5K} (left) The long-time evolution of electron temperature in an ultracold plasma with initial $T_e(0)=5$K in an ellipsoid with a uniform initial density of $0.14 \times 10^{11}$ cm$^{-3}$ (dotted line) and a shell-model Gaussian ellipsoid with initial $\sigma_x = 1.0$ mm, $\sigma_y = 0.55$ mm and $\sigma_z = 0.70$ mm and an initial peak density of $0.4 \times 10^{11}$ cm$^{-3}$ (solid line).  (right) Global population fractions of particles as they evolve in the avalanche of model plasmas formed in an ellipsoid with uniform density (dotted line) and the shell-model ellipsoid (solid line) as above. }
	\end{center}
\end{figure}

Figures \ref{fig:longtime80} and \ref{fig:longtimeT5K} describe the long-time classical evolution of ultracold plasmas with initial conditions chosen to represent possible limits conforming with experimentally measured binding energies and expansion rates of plasmas in states of arrested relaxation \cite{Haenel.2017}. These characteristics suggest a state of Rydberg gas with an average principal quantum number of $n_0=80$, or a plasma of NO$^+$ ions and electrons with a $T_e\le 5$ K.  This ellipsoid has measured Gaussian dimensions of $\sigma_x = 1.0$ mm, $\sigma_y = 0.55$ mm and $\sigma_z = 0.70$ mm.  Analysis of the electron signal over time establishes a peak density at arrest of $0.4 \times 10^{11}$ cm$^{-3}$.  Coupled rate simulations test whether these properties taken as initial conditions describe stable states of a Rydberg gas or ultracold plasma.  Here we see that shell model coupled rate simulations predict evolving populations of charged particles and neutral atoms with electron temperatures rising to exceed 60 K, much the same as classical simulations of corresponding systems with a representative uniform density of $0.14 \times 10^{11}$ cm$^{-3}$.

\subsubsection{Expansion}

Classical hydrodynamics predict that Gaussian ellipsoid plasmas created with initial conditions as described above expand substantially on a timescale of 20 $\mu$s.  Coupled rate-equation simulations for systems of fixed geometry show that details in the distribution of reactants and products do not greatly affect the short-time evolution of electron temperature or global particle densities.  Moreover, as clearly shown in Fig. \ref{fig:shell_dist}, the avalanche spreads very rapidly through the Gaussian ellipsoid on the timescale of expansion. In the inner core, avalanche is complete after $ \SI{10}{\nano\second} $ and from there moves through the plasma until it reaches the wings at $ \SI{1}{\micro\second} $ where most Rydberg molecules have dissociated.  The simulation thus suggests that the classical plasma deviates from a Gaussian ellipsoid only insofar as higher order kinetic rate processes deplete ion densities in the core \cite{Sadeghi.2012}.

These results support the simplification introduced in Section \ref{sect:Methods}, in which coupled-rate equation kinetics of avalanche in individual shells supplies energy globally for an evolution in electron temperature that drives a self-similar expansion of the plasma ellipsoid.  In effect, this assumption neglects distortion of the Gaussian NO$^+$/e$^-$ charged-particle distribution owing to relaxation-rate differentials in the shells, and further implies that the self-similar dynamics of ambipolar expansion extend to describe the expansion of shells in the low-density wings of the plasma where Penning interactions produce very few charged particles.  
\begin{figure}[h!]
\begin{center}
	\includegraphics[width= 0.9 \textwidth]{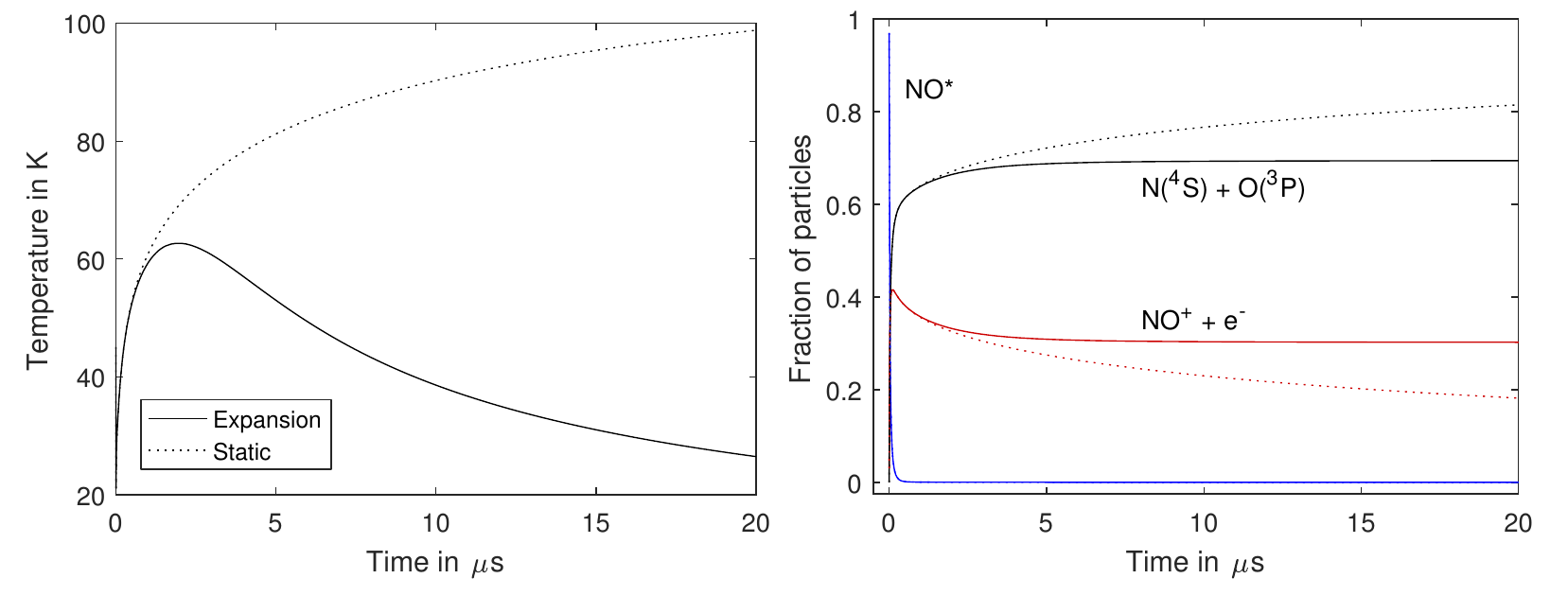}
	\caption{\label{fig:evo49} (left) The long-time evolution of electron temperature in plasmas formed by avalanche from a Rydberg gas with $n_0=49$ in a shell-model Gaussian ellipsoid with initial $\sigma_x = 0.75$ mm, $\sigma_y = \sigma_z = 0.42$ mm and an initial peak density of $0.5 \times 10^{12}$ cm$^{-3}$ for fixed geometry (dotted line) and a stepped sequence of coupled rate equation evolution and self-similar ambipolar expansion (solid line). (right) Global population fractions of particles as they evolve in the avalanche of a shell-model plasma without expansion (dotted line) and with intervening steps of self-similar expansion (solid line) as above. }
	\end{center}
\end{figure}

Figure \ref{fig:evo49} shows how ambipolar expansion affects the evolution of temperature and particle densities for a shell-model Gaussian ellipsoidal plasma initialized as a Rydberg gas with dimensions, $\sigma_x(0) = 0.75$ mm, $\sigma_y(0) = \sigma_z(0) = 0.42$ mm and an initial peak density of $0.5 \times 10^{12}$ cm$^{-3}$.  Here, we see that a transfer of energy from the electron gas to radial motion of the ions reduces $T_e$.  This affects the evolution of particle fractions only to the extent of quenching the conversion of NO$^+$ to N($^4$S) + O($^3$P) owing to the collapse of density that occurs with expansion. 

\begin{figure}[h!]
\begin{center}
	\includegraphics[width= 0.9 \textwidth]{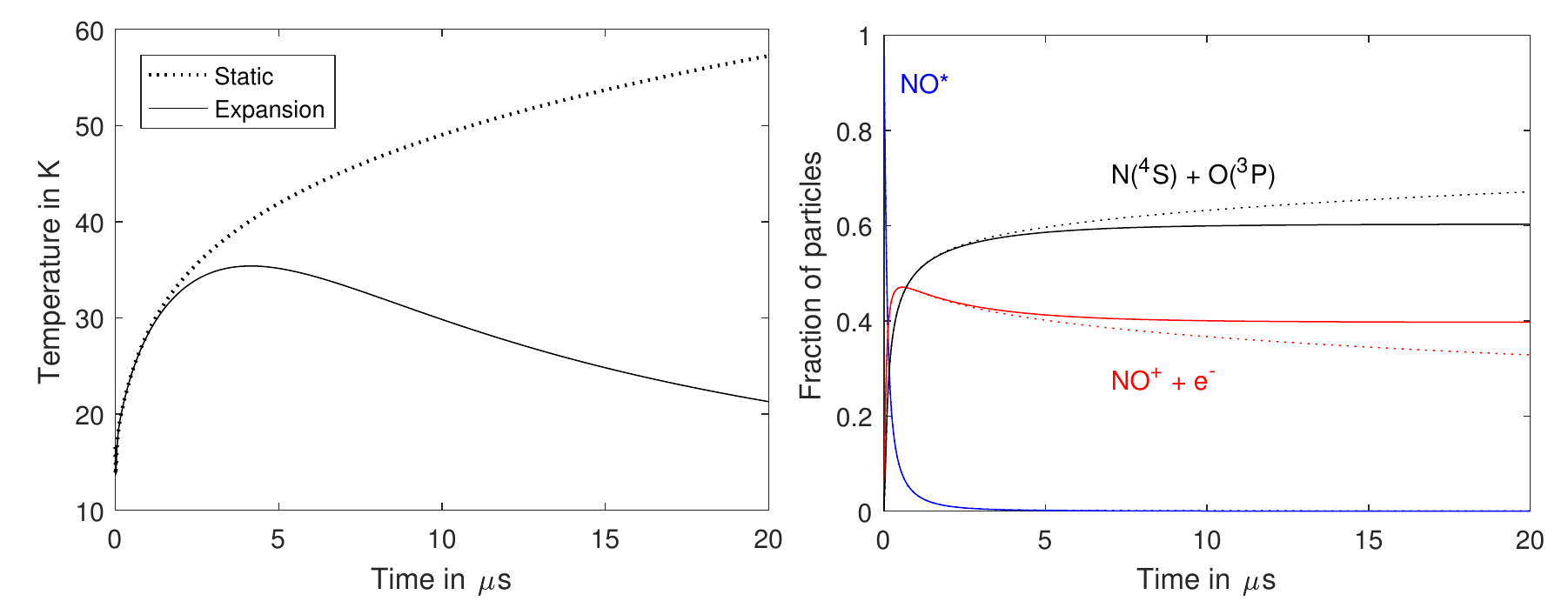}
	\caption{\label{fig:evo80} (left) The long-time evolution of electron temperature in plasmas formed by avalanche from a Rydberg gas with $n_0=80$ in an ellipsoid with a uniform initial density of $0.14 \times 10^{11}$ cm$^{-3}$ (dotted line) and a shell-model Gaussian ellipsoid with initial $\sigma_x = 1.0$ mm, $\sigma_y = 0.55$ mm and $\sigma_z = 0.70$ mm, and an initial peak density of $0.4 \times 10^{11}$ cm$^{-3}$ for fixed geometry (dotted line) and a stepped sequence of coupled rate equation evolution and self-similar ambipolar expansion (solid line). (right) Global population fractions of particles as they evolve in the avalanche of a shell-model plasma without expansion (dotted line) and with intervening steps of self-similar expansion (solid line) as above. }
	\end{center}
\end{figure}

We see a very similar pattern for the test case of a shell-model Gaussian ellipsoidal plasma initialized as a Rydberg gas with dimensions, $\sigma_x = 1.0$ mm, $\sigma_y = 0.55$ mm and $\sigma_z = 0.70$ mm, and an initial peak density of $0.4 \times 10^{11}$ cm$^{-3}$, as shown in Figure \ref{fig:evo80}.  Again, the increasing volume affects the relative fraction of particles only after 5 $\mu$s when falling densities begin to quench electron-ion recombination.

\begin{figure}[h!]
\begin{center}
	\includegraphics[width= 0.9 \textwidth]{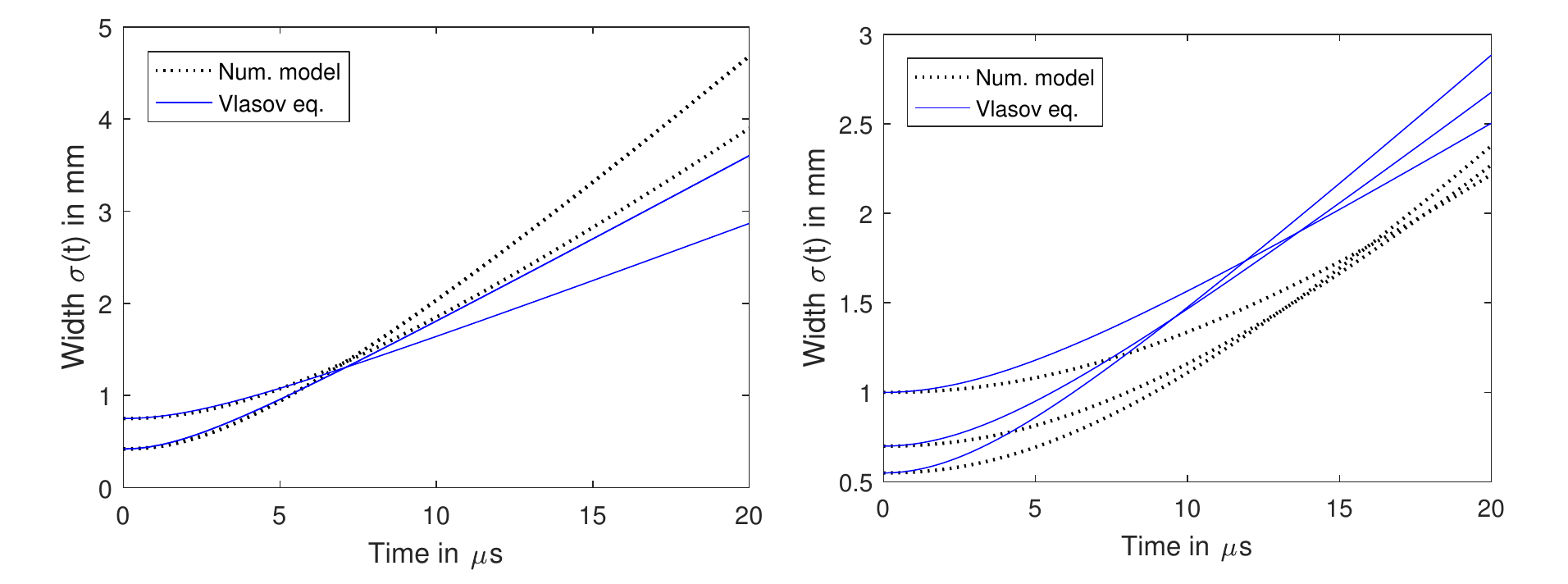}
	\caption{\label{fig:exp} (left) Hydrodynamic expansion along the major axes of Gaussian ellipsoids with initial $\sigma_x = 0.75$ mm, $\sigma_y = \sigma_z = 0.42$ mm and an initial peak density of $0.5 \times 10^{12}$ cm$^{-3}$ approximated after (solid line) avalanche to plasma with $T_e=100$ K (as in Figure \ref{fig:evo49}) by a subsequent self-similar Vlasov expansion of the Gaussian ellipsoid, and (dotted line) concurrent sequential steps of shell-model avalanche and presumed self-similar expansion.  (right) Hydrodynamic expansion along the major axes of Gaussian ellipsoids with initial $\sigma_x = 1.0$ mm, $\sigma_y = 0.55$ mm and $\sigma_z = 0.70$ mm and an initial peak density of $0.4 \times 10^{11}$ cm$^{-3}$ approximated after (solid line) avalanche to plasma with $T_e=60$ K (as in Figure \ref{fig:evo80}) by a subsequent self-similar Vlasov expansion of the Gaussian ellipsoid, and (dotted line) concurrent sequential steps of shell-model avalanche and presumed self-similar expansion. Note ballistic velocities above on the order of $ \SI{150}{\meter\per\second} $.  Simulation results in all cases greatly exceed experimentally measured values of $ 15 $ to $ \SI{30}{\meter\per\second} $.}
	\end{center}
\end{figure}

Following avalanche, coupled rate-equation simulations predict the formation of relatively stable plasmas of NO$^+$ ions and electrons, consistent with experimental observations of arrested decay.  However, this stability relies on a high electron temperature as a means of suppressing ion-electron recombination.  A hot electron gas drives ion expansion.  For an avalanche that proceeds rapidly on the timescale of ion motion, electron heating precedes ambipolar expansion.  We can readily estimate the effect on the size of the ion cloud.  Consider for example the case of stationary ions in an electron gas rapidly heated to an initial temperature, $T_e(0)$.  The electron thermal energy in any single coordinate direction is $\frac{1}{2}k_BT_e(0)$.  Converted entirely to ion kinetic energy this would yield a radial expansion velocity that satisfies $v_i = \sqrt{k_BT_e/m_i}$.  

In the cases illustrated by Figures \ref{fig:longtime49} - \ref{fig:longtimeT5K}, the electron gas gains energy over the full interval of 20 $\mu$s under consideration.  But still, as shown in Figure \ref{fig:exp}, a separable process of electron heating and subsequent self-similar expansion yields about the same ballistic radial velocity as interleaved steps of concurrent avalanche and expansion.  This simply reflects an evolution to quasi equilibrium state in which a roughly equivalent amount of Rydberg binding energy has flowed first to electron thermal motion and from there to the radial acceleration of the ions.  In a concurrent process of avalanche and expansion, this energy flows in steps that proceed before the avalanche is complete in all shells of the plasma.  We can take this insensitivity to path as an indication that the shell model with concurrent expansion paints a reasonable picture of the effect of expansion on the global evolution of state properties in a classical representation of the evolving ultracold plasma.  

\section{Conclusions}

The simulations presented here provide a baseline model from which to proceed in more elaborate efforts to consider the evolution of charge gradients and the ambipolar forces they produce in an experimental system undergoing responsive avalanche dynamics.  We know already that real systems show greater complexity than evidenced by this baseline model.  For example, observations suggest the presence of an important step of energy redistribution driven by resonant charge exchange as radially expanding ions flow from the core of the ellipsoid through quiescent residual Rydberg gas in the wings, in a process that bifurcates the plasma \cite{SchulzWeiling:2016}.  The Rydberg densities necessary to affect this disposition of avalanche energy seem greater than predicted by any classical simulation.  

Allowing for bifurcation accompanied by electron cooling and some degree of recombination to form a population of high-$n$ Rydberg molecules and electron-ion pairs, any coupled rate equation model that conforms with the quenched conditions of density, electron temperature and binding energy observed experimentally evolves to show fast expansion and dissipation that is inconsistent with the state of arrested relaxation observed experimentally.  In contrast with the simulation results above, experimental measures of nitric oxide ultracold plasma expansion consistently produce ballistic expansion velocities in a range from 15 to $ \SI{30}{\meter\per\second} $, far below velocities of 150 m s$^{-1}$ predicted by hydrodynamic rate equation simulations, and these experimentally observed plasma volumes show evident predissociation lifetimes that well exceed 1 ms \cite{Haenel.2017}.  In other work, we have analyzed these departures from well-defined classical expectations to suggest the possibility of quantum effects that arrest the relaxation of the system as it quenches to an ultracold and strongly coupled state \cite{Sous}.   

\section*{Acknowledgements}
 
This work was supported by the US Air Force Office of Scientific Research (Grant No. FA9550-17-1-0343), together with the Natural Sciences and Engineering research Council of Canada (NSERC).

\section*{References}

\bibliography{mybibfile,literature}

\end{document}